\documentclass{elsarticle}

\usepackage{lineno,hyperref}
\modulolinenumbers[5]

\journal{Journal of Computational Physics}

\makeatletter
\def\ps@pprintTitle{%
 \let\@oddhead\@empty
 \let\@evenhead\@empty
 \def\@oddfoot{\vbox{\hsize=\textwidth \footnotesize\itshape
     \vskip 10\p@
     \@journal\hfill Submitted February 12, 2015 \hfill Accepted July 22, 2015 \newline
   \textcopyright~2015, Elsevier. This manuscript version is made available under the CC-BY-NC-ND 4.0 license
   \url{http://creativecommons.org/licenses/by-nc-nd/4.0/}\newline
   doi: \url{10.1016/j.jcp.2015.07.050}\newline
}}%
 \let\@evenfoot\@oddfoot}
\makeatother

\usepackage{textcomp}

\usepackage{longtable}
\usepackage{caption}
\usepackage{amssymb}
\usepackage{amsmath}
\usepackage{url}
\usepackage{psfrag}
\usepackage{color}
\usepackage{etoolbox}
\usepackage[detect-all]{siunitx}
\usepackage{booktabs}
\usepackage{etoolbox}
\usepackage{mathtools}

\usepackage{calc}

\usepackage[letterpaper]{geometry}

\usepackage[normalem]{ulem}

\robustify\itshape 
\robustify\upshape 
\robustify\mdseries 
\robustify\slshape 

\DeclarePairedDelimiter{\norm}{\lVert}{\rVert}

\newcommand{\myeqref}[1]{Eq.~\ref{#1}}  

\newcommand{\HN}{N}
\newcommand{\ark}{a}

\newcommand{\shm}{\mbox{-}}  
\newcommand{\Bpol}{B^{\HN}}
\newcommand{\acoeff}{r}
\newcommand{\secrow}{\slshape}

\newcommand{\sz}[1]{{\scriptsize#1}}
\newcommand{\rmsz}[1]{\mbox{{\sz{\rm{#1}}}}}

\newcommand{\bL}{b\mbox{\tiny$L$}}
\newcommand{\bR}{b\mbox{\tiny$R$}}

\newcommand{\Twall}{T_{\rm WALL}}

\newcommand{\mymat}[1]{{#1}}
\newcommand{\beq}{\begin{equation}}
\newcommand{\eeq}{\end{equation}}
\newcommand{\bec}{\begin{center}}
\newcommand{\eec}{\end{center}}

\newcommand{\plone}[2]{{{\partial} #1}/{\partial {#2}}}
\newcommand{\pltwo}[2]{{{\partial}^2 #1}/{\partial {#2}^2}}

\newcommand{\n}{\noindent}

\newcommand{\TT}{T}

\newcommand{\cmm}{\alpha}
\newcommand{\dmm}{\gamma}

\newcommand{\sq}{\hspace*{0.5 ex}}

\newcommand{\MP}{M}
\newcommand{\LP}{L}

\newcommand{\biprod}{\sigma}

\newcommand{\bo}{\biprod^{m}_{l}}
\newcommand{\bn}[2]{\biprod^{#2}_{#1}}
\newcommand{\cbi}[2]{{{#1} \choose {#2}} }
\newcommand{\wq}[1]{\overline{#1}}
\newcommand{\qq}{W}
\newcommand{\qt}{\wq{W}}
\newcommand{\co}{c^{m}_{j,\,l}}
\newcommand{\cn}[3]{c^{#3}_{#1,\,#2}}
\newcommand{\cc}[3]{c^{m#3}_{j#1,\,l#2}}

\newcommand{\wcheb}[1]{{#1}}

\newcommand{\gauss}[2]{\genfrac{[}{]}{0pt}{}{#1}{#2}_q}
\newcommand{\gaussf}[2]{\left[\genfrac{}{}{0pt}{0}{#1}{#2}\right]_q}

\newcommand{\vph}{\vphantom{\genfrac{}{}{0pt}{0}{P}{k}}}

\newcommand{\FRKC}{\sz{FRKC} }
\newcommand{\FRKCtwo}{\sz{FRKC2} }
\newcommand{\RKC}{\sz{RKC} }
\newcommand{\CVODE}{\sz{CVODE} }
\newcommand{\SERK}{\sz{SERK} }

\newcommand{\ROCKtwo}{\sz{ROCK2} }
\newcommand{\ROCKf}{\sz{ROCK4} }
\newcommand{\DUMKA}{\sz{DUMKA} }

\newcommand{\pFRKC}{\sz{FRKC}}
\newcommand{\pFRKCtwo}{\sz{FRKC2}}
\newcommand{\pRKC}{\sz{RKC}}
\newcommand{\pCVODE}{\sz{CVODE}}

\newcommand{\pROCKf}{\sz{ROCK4}}

\newcommand{\CVODEtwo}{\sz{CVODE2} }
\newcommand{\pCVODEtwo}{\sz{CVODE2}}

\newcommand{\FRKCs}{\sz{FRKCs} }
\newcommand{\pFRKCs}{\sz{FRKCs}}
\newcommand{\FRKCtwos}{\sz{FRKC2s} }

\newcommand{\FRKCfs}{\sz{FRKC4s} }
\newcommand{\pFRKCfs}{\sz{FRKC4s}}
\newcommand{\FRKCxs}{\sz{FRKC6s} }
\newcommand{\pFRKCxs}{\sz{FRKC6s}}

\newcommand{\aspac}{\mbox{\hspace*{3.6ex}}}

\begin{document}

\begin{frontmatter}

\title{A class of high-order Runge-Kutta-Chebyshev stability polynomials}
\author{Stephen O'Sullivan}
\address{School of Mathematical Sciences, Dublin Institute of Technology, Kevin Street, Dublin 8, Ireland}
\ead{stephen.osullivan@dit.ie}

\begin{abstract}
The analytic form of a new class of factorized Runge-Kutta-Chebyshev (\pFRKC) stability polynomials of arbitrary order $N$ is presented. Roots of \FRKC stability polynomials of degree $L=MN$ are used to construct explicit schemes comprising $L$ forward Euler stages with internal stability ensured through a sequencing algorithm which limits the internal amplification factors to $\sim L^2$.   The associated stability domain scales as $M^2$ along the real axis. Marginally stable real-valued points on the interior of the stability domain are removed via a prescribed damping procedure.

By construction, \FRKC schemes meet all linear order conditions; for nonlinear problems at orders above 2, complex splitting or Butcher series composition methods are required. Linear order conditions of the \FRKC stability polynomials are verified at orders 2, 4, and 6 in numerical experiments.  Comparative studies with existing methods show the second-order unsplit \FRKCtwo scheme and higher order (4 and 6) split \FRKCs schemes are efficient for large moderately stiff  problems.\\

\end{abstract}


\end{frontmatter}

 
\section{Introduction}
\label{intro}

Runge-Kutta-Chebyshev methods are explicit numerical integration schemes with extended stability domains derived from the optimality properties of Chebyshev polynomials~\citep{markov1890question, markov1892functions}. These methods are commonly applied to moderately stiff  systems of semi-discrete equations of the form
\beq
w'=f(t,\,w) ,
\label{eqn:firsteqn}
\eeq
yielding an approximate solution $w^n$ at time $t^n=n \TT$, defined on a spatial mesh of spacing $h$ at points $x_k$, with $x_{k+1}=x_{k}+h$. Such systems arise naturally through application of the method of lines to parabolic systems. Runge-Kutta-Chebyshev methods may be broadly categorized as factorized or recursive in nature.

Factorized Runge-Kutta-Chebyshev methods are formed from a sequence of forward Euler stages. These methods were first suggested by \citet{saul1960integration,guillou1960domaine} and were subsequently considered by \citet{gentzsch1978one} and \citet{vanderHouwen1996261}. They have been applied at first-order and extended to second-order via Richardson extrapolation by various authors~\citep{alexiades1996super,o2006explicit,o2007three,o2011acceleration}. Based on a strategy proposed by \citet{lebedev2000explicit}, the \DUMKA stability polynomials exist at orders 2, 3, and 4~\citep{medovikov1998high}.

Recursive Runge-Kutta-Chebyshev methods were first described by \citet{vanderhouwen1980} and rely on (three-term) recursion to generate a solution. They were introduced at second-order by \citet{vanderhouwen1980} and, subsequently, other second, third, and fourth-order methods have been developed~\citep{verwer1996explicit,sommeijer1998rkc,abdulle2001second,abdulle2002fourth,martin2009second}. We note that alternative approaches with second-order accuracy involving Legendre polynomials have recently been proposed by \citet{meyer2012second,meyer2014stabilized}. At orders above 2, for both factorized and recursive methods, composition techniques relying on Butcher series theory \citep{hairernonstiff,zbMATH05292586} are typically used to satisfy the full set of order conditions~\citep{medovikov1998high,abdulle2002fourth}.

This paper is organized as follows. In Section~\ref{sec:frkc}, the analytic form of the class of \FRKC  stability polynomials is presented. The construction of stable time-marching schemes based on the roots of these polynomials is outlined. Section~\ref{sec:frkcderiv} is given over to the derivation of the polynomial through consideration of associated recurrence relations. 
In Section~\ref{sec:applic}, numerical tests are presented confirming the order and efficiency properties of \FRKC methods. Conclusions are presented in Section~\ref{sec:conc}.

\section{High-order factorized Runge-Kutta-Chebyshev}
\label{sec:frkc}

\subsection{General prescription}

Eq.~\ref{eqn:firsteqn} may be written in autonomous form by appending $t$ to the vector of dependent variables for the system
\beq
w'=f(w) . 
\label{eqn:autoneqn} 
\eeq
Parentheses may be used in the remainder of this work to differentiate exponents from indices. We proceed by considering order $N$ extended stability explicit Runge-Kutta schemes over $L= MN$ stages

\beq
\mymat{W}^{L}=\mymat{W}^0+\TT\sum^{L}_{l=1}\ark_l f(W^{l-1}) ,
\label{eqn:rk}
\eeq

\n where $\mymat{W}^0=w^n$ corresponds to the approximate solution $\mymat{w}^{n}$ at time level $n$, and $\mymat{W}^{L}$ yields $\mymat{w}^{n+1}$ at a time $\TT$ later. The timestep related to each stage is then given by $\tau_l= \ark_l\TT$.

The \FRKC polynomial of rank $N$, and degree $L$, is given by

\beq
\label{eqn:FRKC}
B^{\HN}_M(z) = d^{\HN}_0 + 2\sum_{k=1}^{\HN} d^{\HN}_k C_{kM}(z) ,
\eeq

\n where $C_{kM}$ denotes the the Chebyshev polynomial of the first kind of degree $kM$. The corresponding optimal real stability range is $[-\beta_M,\,0]$, where $\beta_M=2M^2\cmm_M$, $\cmm_M= (\dmm_M N+2)/3$, and $\dmm_2\approx 0.87$ (with $\dmm_M$ rapidly converging to 1 as $M$ increases). In this limit, the polynomials generate  $81\%$, $74\%$ and $73\%$ of the optimal intervals for order $2,\,4,\,6$ respectively~(see \citet{van1977} and \citet{abdulle2001chebyshev} for estimates of the optimal values for $\cmm_M$).  The limiting step size is $\TT=\beta_M/|\lambda|_{\rmsz{max}}$, where $\lambda$ are the negative-definite eigenvalues for the Jacobian of Eq.~\ref{eqn:autoneqn}. We note that the form of \myeqref{eqn:FRKC} is consistent with the known result that Chebyshev expansions of stability polynomials exist to arbitrary order~\citep{bakker71}. Furthermore, following from a proposition by \citet{lomax1968construction}, \citet{riha72} confirmed the existence and uniqueness of optimal stability polynomials with $L-N$ local maxima with value unity. A full derivation of the \FRKC polynomial expression given by Eq.~\ref{eqn:FRKC} is provided in Section~\ref{sec:frkcderiv}.  

%
\begin{figure*}
  \centering
  \begin{minipage}{160mm}
    \centering\leavevmode
    {
      \hspace*{-5mm}
      \resizebox{79mm}{!}{\input{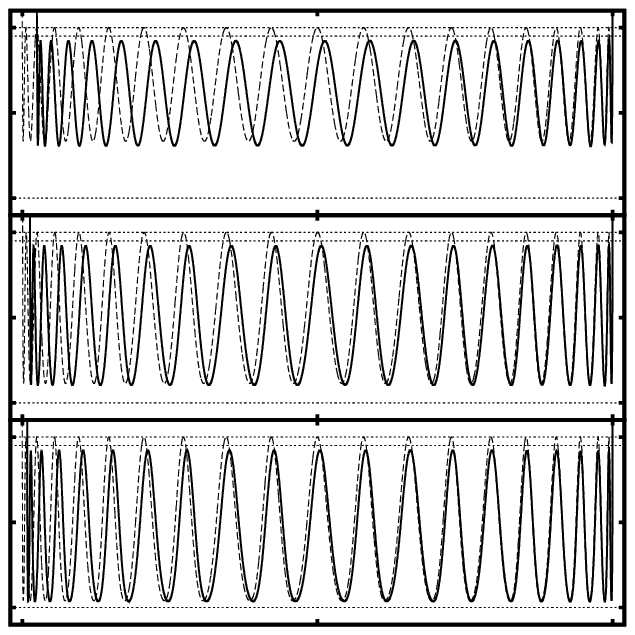}}
      \hspace*{-5mm}
      \resizebox{79mm}{!}{\input{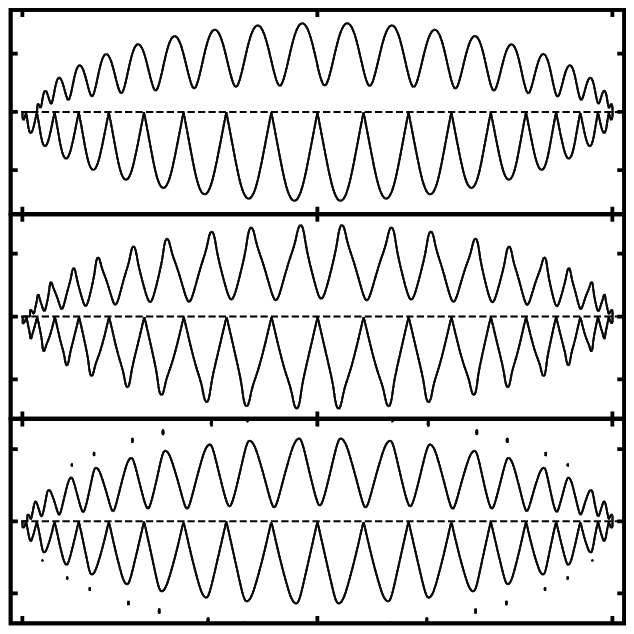}}
    }
\end{minipage}
  \caption{Absolute values along the real axis for \FRKC stability polynomials corresponding to $M=20$ at various values of $\HN$. Damped polynomials with $\nu_0=0.05$  are shown with solid lines in the left side panels, and for $y>0$ in the right side panels; associated undamped polynomials are also shown with dashed lines on the left, and for $y\le 0$ on the right. For $N=2$: $\dmm_{20}=0.9988$, $\beta_{20}=1066.0$; for $N=4$: $\dmm_{20}=1.0215$, $\beta_{20}=1623.9$; for $N=6$: $\dmm_{20}=1.0276$, $\beta_{20}=2177.5$. Left side panels show plots of $|R^{\HN}_M(x)|$ ($x \in \mathbb{R}$): \textbf{(a)} $|R^{2}_{20}(x)|$; \textbf{(b)} $|R^{4}_{20}(x)|$; \textbf{(c)} $|R^{6}_{20}(x)|$. Dotted lines indicate guide values at 1.0, 0.95, 0.0. Right side panels show $|R^{\HN}_M|=1$: \textbf{(d)} $|R^{4}_{20}|=1$; \textbf{(e)} $|R^{4}_{20}|=1$; \textbf{(f)} $|R^{6}_{20}|=1$. 
  }
  \label{fig:1}       
\end{figure*}

The order coefficients $d^{\HN}_k$, which we refer to collectively as the order \emph{pattern}, are determined by requiring that the (undamped) stability polynomial $R^{\HN}_M(z)=B^{\HN}_M(1+z/\MP^2\cmm_M)$, consisting of shifted Chebyshev polynomials, satisfies the linear order conditions 
\beq
R^{\HN}_M{}^{(n)}(0)=1 , \quad n=1,\,\cdots,\,N .
\label{eqn:linearorderconds}
\eeq 
This requirement is met by solving the $N$-dimensional linear system\footnote{The identity $C_{kM}^{(l)}(1)=\prod_{i=0}^{l-1}((kM)^2-i^2)/(2i+1)$ is useful here.} 

\beq
\label{eqn-pattern}
\left[ \begin{array}{ccc}
C_{M}^{(1)}(1) & \ldots & C_{NM}^{(1)}(1) \\
\vdots    & \ddots & \vdots \\
C_{M}^{(N)}(1) & \ldots & C_{NM}^{(N)}(1) 
\end{array}\right]
\left[\begin{array}{c}
d^{\HN}_1\\
\vdots\\
d^{\HN}_N
\end{array}\right]
=
\left[\begin{array}{c}
(\MP^2\cmm_M)^1\\
\vdots\\
(\MP^2\cmm_M)^N
\end{array}\right] ,
\eeq

\n coupled with the constraint

\beq
d^{\HN}_0=1-2\sum_{k=1}^{\HN} d^{\HN}_k .
\eeq
\n Following identification of the roots $\zeta_l$ of the \FRKC polynomial $B^{\HN}_M(z)$, the damped order $N$ scheme corresponding to Eq.~\ref{eqn:rk} is determined via
\beq
\ark_l=\frac{1}{\MP^2\cmm_M}~\frac{1}{1-\zeta_l} .
\label{eqn-taunodamp}
\eeq
In order to ensure a stable scheme for small perturbations from the real axis in the spectrum of Eq.~\ref{eqn:autoneqn}, it is necessary to introduce a suitable damping procedure. We find an effective prescription for the damped order $N$ scheme is given by
\beq
\ark_l=\frac{1}{(1-\nu)\MP^2\cmm_M}~\frac{1-\mu_l}{1-(1-2\mu_l)\zeta_l} ,
\label{eqn-tau}
\eeq
\n where the damping is parameterized by the small positive quantity $\nu$, resulting in the real extent of the stability interval being reduced to $(1-\nu)\beta_M$. The value of $\nu=\nu_0/N$ is regulated by means of the reference damping parameter $\nu_0$, such that maxima in $|R|$ along the real axis are scaled by approximately $1-\nu_0$.

For the case $\nu_0=0.05$, with $M=20$, and for various values of $N$, Fig.~\ref{fig:1} illustrates the effect of the damping procedure. It is clear that the undamped polynomials are marginally stable at $M-1$ points on the interior of the stability domain along the real axis. (In fact, for sub-optimal $\cmm_M$, internal marginally stable points occur at $M/2-1$ locations for even values of $M$, or $(M-1)/2$ locations for odd values of $M$.) Examples of the order patterns for $M=20$ with $N=2,\,4,\,6$ are given in \ref{app:patterns}.

The $L$-tuple $[\mu_l]$ has cardinality $\HN$ and regulates the implementation of damping in the scheme while preserving the nominal order of accuracy. The values of $\mu_l$ are obtained by tuning the damped stability polynomial $R^{\HN}_M(z)=\Pi_{l=1}^L(1+\ark_l z)$ to meet the linear order conditions given in Eq.~\ref{eqn:linearorderconds}. We describe the procedure for the determination of the damping coefficients $\mu_l$ in Section~\ref{sec:damp}. 

\subsection{Identification of damping parameter $L$-tuple}
\label{sec:damp}

The elementary symmetric polynomial, $\bn{l}{m}= \sum_{1 \le j_1 < \dots < j_l \le m} \prod_{i=1}^{l} \zeta_{j_i}$, is defined as the sum of all possible products formed from $l$ unrepeated elements drawn from the first $m$ elements of an $L$-tuple $[\zeta_l]$. The definition is extended by setting $\bn{0}{m}=1$ and $\bn{k>m}{m}=0$. We associate the $L$ roots $\zeta_l$, in order of increasing real component $\Re(\zeta_l)$, with the damping coefficients $\mu_l$ by cycling through the $N$ damping coefficients a total of $M$ times. Newton-Raphson iterations then converge rapidly to the linear order conditions given by Eq.~\ref{eqn:linearorderconds}. The effects of the damping procedure on the stability domain are shown in Fig.~\ref{fig:1}. The stage intervals $\tau_l$ given by Eq.~\ref{eqn-tau} are complex in general, however, with $d^{1}_0=0$, $d^{1}_1=1/2$, the standard first-order super-timestepping scheme~\citep{alexiades1996super,o2006explicit} is recovered with $B^{1}_M = C_{M}$. For $N>1$, either one or two values of $\tau_{l}$ have negative real parts.

The presented prescription implements conjugate pairs separately thereby necessitating full complex arithmetic. Other than some penalty in the additional computational demand required, we find no practical disadvantage to preserving this model of treating each factor as distinct. We note that \citet{lebedev2000explicit,lebedev1994solve} proposed a scheme which treats stages in pairs and, when applied to conjugate pairs, removes the need for complex arithmetic.

\begin{figure*}
  \centering
  \begin{minipage}{160mm}
    \centering\leavevmode
      {
      \resizebox{160mm}{!}{\input{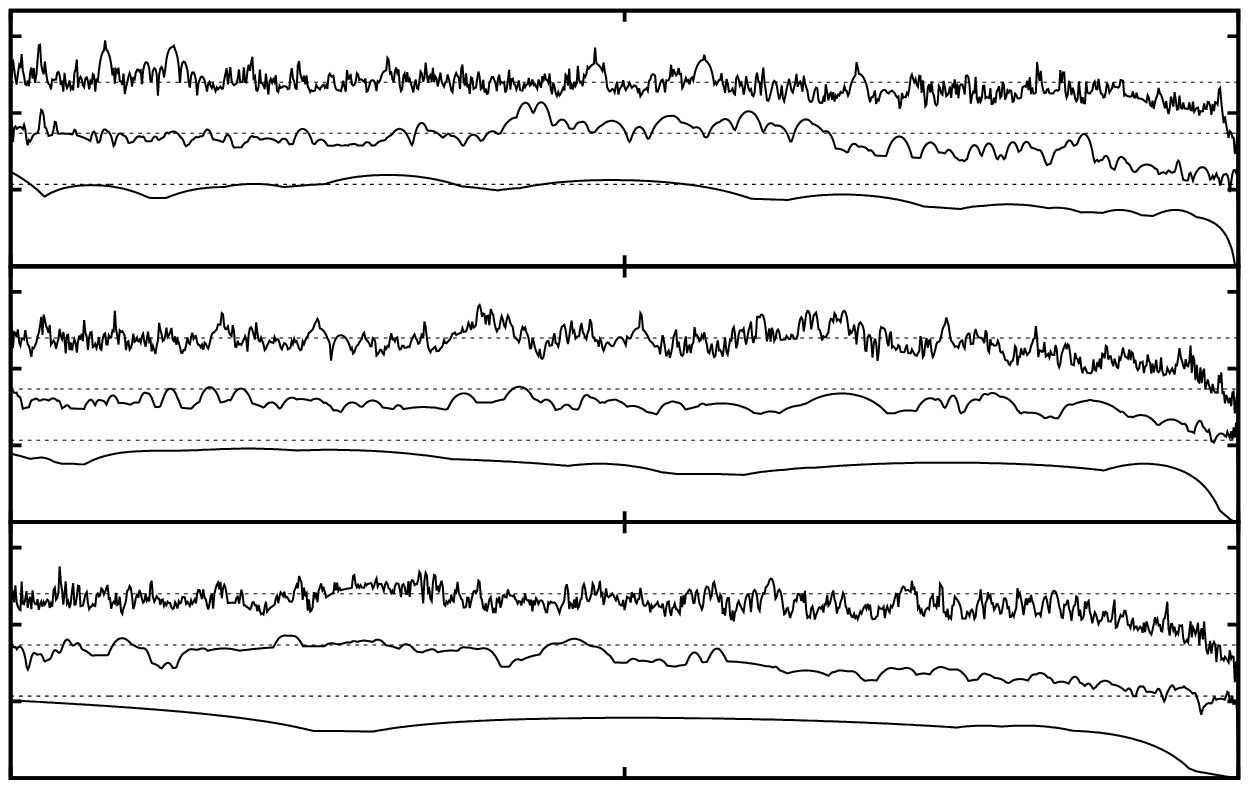}}
      }
  \end{minipage}
  \caption{Maximum internal stability function $Q(x)$ for $L\approx 4000$ (upper lines), $L\approx 400$ (middle lines), $L\approx 40$ (lower lines): \textbf{(a)} $N=2$ with $M=2000,\,200,\, 20$; \textbf{(b)} $N=4$ with $M=1000,\,100,\,10$; \textbf{(c)} $N=6$ with $M=667,\,67,\,7$. In all cases the default value of $\nu_0=0.05$ is used. Guidelines show values of $\LP^2$.}
  \label{fig:internal}       
\end{figure*}

\subsection{Internal stability}
\label{sec:intstab}

Schemes comprising a high number of stages are internally unstable if the sequencing of the stages is allowed to admit uncontrolled growth of numerical errors \citep{van1977,lebedev1976utilization, verwer1990convergence,  hairerstiff, ketcheson2013propagation}. \citet{lebedev1973solution} first suggested sequencing stages to manage uncontrolled growth of internal instabilities (see also \citep{marchuk1986numerical}). Here, we present a straightforward algorithm for sequencing stages which limits the maximum amplification factor of internal instabilities to $\sim L^2$.

We define $v_{j,\,k}= |1+\ark_j x_k|$, where $x_k\in [-\beta_M,\,0]$ are discrete values spanning the spectrum of Eq.~\ref{eqn-scheme1}. The $L$-tuple $[\ark_l]$ is then ordered by holding the ${\rm L}_1$ normed quantity
\beq
\norm*{\max\left(\prod_{j=1}^{l}v_{j,\,k},\,\prod_{j=l+1}^{L}v_{j,\,k}\right)}_1
\eeq
to a minimum value while $l$ is increased from 1 to $L$. This procedure suppresses the growth of the internal stability functions $Q_{j,\,k}(x)=\prod_{l=j}^k|1+\ark_l x|$, for $j,\,k=1,\,\cdots,\,L$, over $x\in[-\beta_M,\,0]$, and provides excellent internal stability properties with high numbers of stages at low computational cost. 
In Fig.~\ref{fig:internal}, we plot the maximum internal stability function $Q(x)=\max_{j,\,k}(Q_{j,\,k}(x))$ for the test cases $L\approx 4000,\,400,\,40$, with $N=2,\,4,\,6$, and $\nu_0=0.05$. The optimization may be enhanced by concentrating the points $x_k$ towards the bounds of the interval. (In this work a logistic function over a range of 15 is employed to generate the $L$ sample points.) We observe the maximum internal amplification factor scales approximately as $L^2$. Hence, the internal stability properties are well within the acceptable limits of modern computing precision for any practical problem.

Consistent with these findings, we note that internal amplification factors of $\sim 10^6$ are quoted in the literature for \RKC methods with $1000$ stages~\citep{hundsdorfer2003numerical}, and furthermore, a quadratic dependence on stage number is suggested by \citet{sommeijer1998rkc}. \ROCKtwo methods are reported to demonstrate amplification factors of $\sim 10^9$ at 200 stages by~\citet{hundsdorfer2003numerical}, suggesting internal instability growth rates 150 time larger than for \RKC and \FRKCtwo schemes.

We note that the \SERK scheme is also limited in stage number, requiring 600 digits of precision for 320 stages, albeit principally due to severely ill-conditioned matrix systems used in calculating the stability polynomials by means of the Remez algorithm~\citep{martin2009second}. A subsequent revision of the \SERK methodology has demonstrated a stability range which is four times larger~\cite{martin2014stabilized}.

\section{Factorized Runge-Kutta-Chebyshev polynomial derivation}
\label{sec:frkcderiv}

We proceed by considering the one-dimensional diffusion equation

\beq
\frac{\partial w}{\partial t}=\frac{\partial^2 w}{\partial x^2} .
\label{eqn:1dheat}
\eeq
\n The semi-discrete form of \myeqref{eqn:1dheat} may be written $\mymat{w}'=h^{-2}\mymat{D}\mymat{w}$, where $\mymat{D}$ is a tridiagonal matrix with diagonal entries -2, subdiagonal entries 1, corresponding to a second-order central discretization of the spatial derivative. The eigenvalues of $\mymat{D}$ are negative with a maximum magnitude of 4. Application of the numerical scheme given by Eq.~\ref{eqn:rk} yields
\beq
\mymat{w}^{n+1}=\prod^L_{l=1}\left(\mymat{I}+\frac{\tau_l}{h^2}\mymat{D}\right)\mymat{w}^n .
\label{eqn-scheme1}
\eeq

\begin{figure*}[!ht]
  \psfrag{ppp1}{\hspace*{0.2mm}\vspace*{0.1mm}\scriptsize \textbf{\phantom{-}1}}
  \psfrag{ppp2}{\hspace*{0.2mm}\vspace*{0.1mm}\scriptsize \textbf{\phantom{-}2}}
  \psfrag{ppm1}{\hspace*{0.2mm}\vspace*{0.1mm}\scriptsize \textbf{-1}}
  \psfrag{ppm2}{\hspace*{0.2mm}\vspace*{0.1mm}\scriptsize \textbf{-2}}

  \psfrag{ppatit}{\scriptsize\hspace*{0.0mm}\vspace*{0.0mm} $\mathbf{\acoeff^{5}_0\aspac \acoeff^{5}_1\aspac \acoeff^{5}_2\aspac \acoeff^{5}_3}$}
  \psfrag{ppa0}{}
  \psfrag{ppa1}{\hspace*{0.0mm}\vspace*{0.0mm}$\mathbf{\shm1\quad\phantom{\shm}1\quad\phantom{\shm}1}$}
  \psfrag{ppa2}{\hspace*{0.0mm}\vspace*{0.0mm}$\mathbf{\phantom{\shm}0\quad \shm1\quad\phantom{\shm}1\quad\phantom{\shm}1}$}
  \psfrag{ppa3}{\hspace*{0.0mm}\vspace*{0.0mm}$\mathbf{\phantom{\shm}0\quad \shm1\quad\phantom{\shm}1\quad\phantom{\shm}1}$}
  \psfrag{ppa4}{\hspace*{0.0mm}\vspace*{0.0mm}$\mathbf{\shm1\quad\phantom{\shm}1\quad\phantom{\shm}1}$}
  \psfrag{ppa5}{\hspace*{0.0mm}\vspace*{0.0mm}$\mathbf{\phantom{\shm}1}$}

  \psfrag{ppg1}{\hspace*{0.0mm}\vspace*{0.0mm}$\mathbf{\phantom{\shm}\gaussf{5}{1}}$}
  \psfrag{ppg2}{\hspace*{0.0mm}\vspace*{0.0mm}$\mathbf{         \shm \gaussf{5}{2}}$}
  \psfrag{ppg3}{\hspace*{0.0mm}\vspace*{0.0mm}$\mathbf{\phantom{\shm}\gaussf{5}{3}}$}
  \psfrag{ppg4}{\hspace*{0.0mm}\vspace*{0.0mm}$\mathbf{         \shm \gaussf{5}{4}}$}
  \psfrag{ppg5}{\hspace*{0.0mm}\vspace*{0.0mm}$\mathbf{\phantom{\shm}\gaussf{5}{5}}$}

  \psfrag{ppa}{\hspace*{0.0mm}\vspace*{0.0mm}$\mathbf{(a)}$}
  \psfrag{ppb}{\hspace*{0.0mm}\vspace*{0.0mm}$\mathbf{(b)}$}
  \psfrag{ppc}{\hspace*{0.0mm}\vspace*{0.0mm}$\mathbf{(c)}$}

  \psfrag{ppmmtit}{$\mathbf{m}$}
  \psfrag{ppmm0}{\hspace*{0.0mm}\vspace*{0.0mm}$\mathbf{m=0}$}
  \psfrag{ppmm1}{\hspace*{0.0mm}\vspace*{0.0mm}$\mathbf{1}$}
  \psfrag{ppmm2}{\hspace*{0.0mm}\vspace*{0.0mm}$\mathbf{2}$}
  \psfrag{ppmm3}{\hspace*{0.0mm}\vspace*{0.0mm}$\mathbf{M-5}$}
  \psfrag{ppmm4}{\hspace*{0.0mm}\vspace*{0.0mm}$\mathbf{M-4}$}
  \psfrag{ppmm5}{\hspace*{0.0mm}\vspace*{0.0mm}$\mathbf{M-3}$}
  \psfrag{ppmm6}{\hspace*{0.0mm}\vspace*{0.0mm}$\mathbf{M-2}$}
  \psfrag{ppmm7}{\hspace*{0.0mm}\vspace*{0.0mm}$\mathbf{M-1}$}
  \psfrag{ppmm8}{\hspace*{0.0mm}\vspace*{0.0mm}$\mathbf{M}$}

  \psfrag{ppltit}{$\mathbf{k}$}
  \psfrag{ppl0}{\hspace*{0.0mm}\vspace*{0.0mm}$\mathbf{}$}
  \psfrag{ppl1}{\hspace*{0.0mm}\vspace*{0.0mm}$\mathbf{1}$}
  \psfrag{ppl2}{\hspace*{0.0mm}\vspace*{0.0mm}$\mathbf{2}$}
  \psfrag{ppl3}{\hspace*{0.0mm}\vspace*{0.0mm}$\mathbf{3}$}
  \psfrag{ppl4}{\hspace*{0.0mm}\vspace*{0.0mm}$\mathbf{4}$}
  \psfrag{ppl5}{\hspace*{0.0mm}\vspace*{0.0mm}$\mathbf{5}$}

  \psfrag{ppjtit}{}
  \psfrag{ppjm3}{\hspace*{-1.0mm}\vspace*{0.0mm}\scriptsize \textbf{$\mathbf{j}$-3}}
  \psfrag{ppjm2}{\hspace*{-1.0mm}\vspace*{0.0mm}\scriptsize \textbf{$\mathbf{j}$-2}}
  \psfrag{ppjm1}{\hspace*{-1.0mm}\vspace*{0.0mm}\scriptsize \textbf{$\mathbf{j}$-1}}
  \psfrag{ppj0}{\hspace*{0.5mm}\vspace*{0.0mm}\scriptsize \textbf{$\mathbf{j}$}}
  \psfrag{ppjp1}{\hspace*{-1.5mm}\vspace*{0.0mm}\scriptsize \textbf{$\mathbf{j}$+1}}
  \psfrag{ppjp2}{\hspace*{-1.0mm}\vspace*{0.0mm}\scriptsize \textbf{$\mathbf{j}$+2}}
  \psfrag{ppjp3}{\hspace*{-1.0mm}\vspace*{0.0mm}\scriptsize \textbf{$\mathbf{j}$+3}}

  \includegraphics[height=12cm,angle=0]{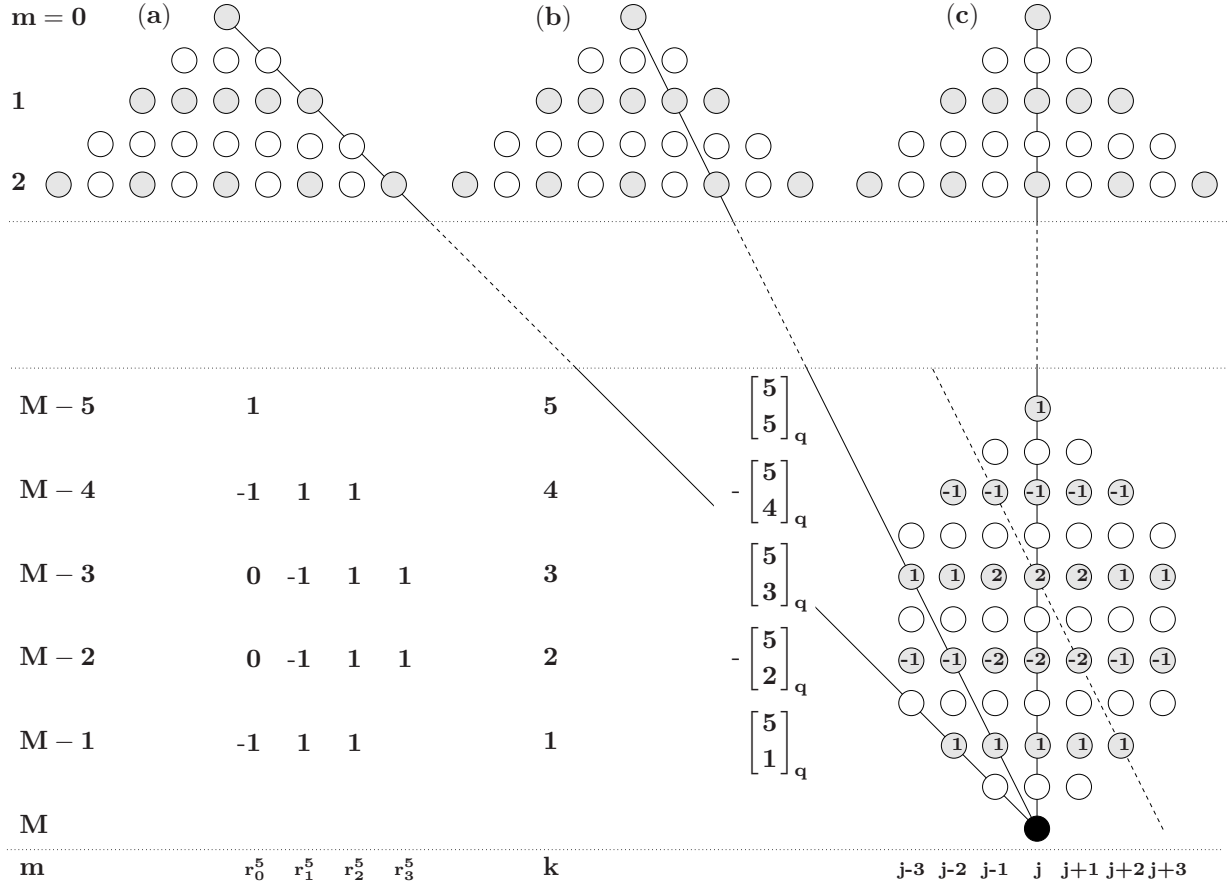}
  \caption{Graphical representation of construction of primitive recurrence relation between state value $\qt^{\wcheb{L}}_j$ and solution states for $m<L$ at intervals of $\HN=2$; non-zero coefficients used in~\myeqref{eqn:primrecur} corresponding to Gaussian polynomials $\gauss{P}{k}$ are shown ($P=2N-1$). Nodes at values of $m$ which are non-integral multiples of $\HN$ (empty circles) do not appear in the relation construction.
Pattern flows emerging from three sample source distributions up to segment level $m=2$ are shown  (labeled \emph{a}, \emph{b}, and \emph{c}). Rays terminating from the filled node at $m=L$ and originating at the apices of the sample distributions  are shown summing to unity (solid lines). Rays which do not similarly project the solution pattern from $m=2$ through to $m=L$ sum to zero (dashed line).  Also shown are the coefficients $\acoeff^{P,\,k}_g$ prescribed by~\myeqref{eqn:acoeffs} at given values of $k$.}
  \label{fig:recur}
\end{figure*}

The \FRKC polynomial $B^{\HN}_M$ may be derived by consideration of the canonical scheme given by~\myeqref{eqn-scheme1} over an extended timestep $\TT$, spanning time levels $t^n$ to $t^{n+1}$, and consisting of $M$ \emph{segments}, with each segment comprising $N$ stages. We write the solution state corresponding to $\mymat{w}^{n}$ as $\mymat{W}^0$, and assume $W^0_0=1$ and $W^0_{k\ne 0}=0$; more complex states may be constructed by superposition. The solution state corresponding to $\mymat{w}^{n+1}$ is then obtained from $\mymat{W}^{M}=\prod^{MN}_{l=1}(\mymat{I}+h^{-2}\tau^M_l\mymat{D})\mymat{W}^0$. To aid the following discussion, Fig.~\ref{fig:recur} is provided to graphically represent solution states $\mymat{W}^m$ at different segment levels for the particular case $N=2$. A reference point value of the solution state $\mymat{W}^{M}$, at spatial index $j$, is shown as a black node.

To proceed, we assume schemes consisting of $m$ segments ($m=1,\,\cdots,\,M-1$) are known which generate the solution states, $\mymat{W}^{m}=\prod^{m\HN}_{l=1}(\mymat{I}+h^{-2}\tau^m_l\mymat{D})\mymat{W}^0$. For $m=1$, the solution state $\mymat{W}^{1}$ spans $2N+1$ nodes from a given point profile $\mymat{W}^0$. Successive states regenerate this pattern, but spanning $(2mN+1)$ nodes, with non-zero values interspersed by $(m/\HN-1)$ zero valued nodes. We refer to the sequence of patterns over increasing values of $m$ as a pattern \emph{flow}. Illustrations of sample pattern flows are given in Fig.~\ref{fig:recur}.

Using Eq.~\ref{eqn-taunodamp}, the components of the states $\mymat{W}^m$ may be recast as $\qt^m_j=\qq^m_j\prod^{m\HN}_{l=1}(1-\zeta^m_l)$. Over a single timestep, \myeqref{eqn-scheme1} then takes the simplified form

\beq
\wq{\mymat{W}}^{m}=\prod^{m\HN}_{l=1}\wq{\mymat{D}}^m_l\wq{\mymat{W}}^0 ,
\label{eqn-scheme2}
\eeq
\n where $\wq{\mymat{D}}^m_l$ is a tridiagonal matrix with diagonal entries $-\zeta^m_l$ and subdiagonal entries 1/2. In terms of the elementary symmetric polynomials we have
\beq
\label{eqn:coeffs}
\qt^{m}_j= \sum^m_{l=0}\co\bo ,
\eeq where $\co$ are coefficients dependent on the scheme~\myeqref{eqn-scheme2}. By induction, these coefficients have the properties

\begin{equation}
\label{eqn:co}
\begin{aligned}[c]
&\cn{0}{m}{m} &=& (-1)^m ,\\ 
&\cn{j}{l\ne m}{m} &=& \frac{1}{2}\left(\cc{-1}{}{-1}+\cc{+1}{}{-1}\right) ,
\end{aligned}
\quad\quad
\begin{aligned}[c]
&\cn{j\ne 0}{m}{m} &=& 0 ,\\
&\cn{j}{l\ne 0}{m}  &=& -\cc{}{-1}{-1} .
\end{aligned}
\end{equation}
\newpage

\newgeometry{left=25mm,right=25mm}
\restoregeometry

\n The ($m+1$)-tuples $[\bn{l}{m}]$ fully determine $\wq{\mymat{W}}^m$ through the roots $\zeta^{m}_l$ of the associated polynomial $\Bpol_{m}$ defined by
\beq
\label{eqn:poly}
\frac{\Bpol_{m}}{(2)^{m\HN-1}}=\sum_{l=0}^{m}(-1)^l \bn{l}{m} (z)^{m-l} ,
\eeq 
\n where $(2)^{m\HN-1}$ is a normalization factor. Hence, the $(mN)$-tuple $[\tau^{m}_l]$ is completely specified by $[\bn{l}{m}]$.

\subsection{Derivation}

The ($m+1$)-tuple $[0^{\bL},\,\bn{}{m-b},\,0^{\bR}]$ is constructed from the elementary symmetric polynomials corresponding to the solution $\wq{\mymat{W}}^{m-b}$, where $\bn{}{m-b}$ indicates the ordered elements $\bn{0}{m-b},\,\ldots,\,\bn{m-b}{m-b}$, zero superscripts denote multiplicity, and $b= \bL+\bR$. Through \myeqref{eqn:poly}, the ($m+1$)-tuple maps to the degree $m-\bL$ polynomial $(-1)^{\bL}(z)^{\bR} (2)^{-(m-b)\HN+1}\Bpol_{m-b}$. Inserting $[0^{\bL},\,\bn{}{m-b},\,0^{\bR}]$ into \myeqref{eqn:coeffs} and appealing to the properties of the coefficients $\co$, as given by \myeqref{eqn:co}, yields a direct correspondence to 
$\left(-1\right)^{\bL}\left(\frac{1}{2}\right)^{\bR} \sum_{g=0}^{\bR}\cbi{\bR}{g} \qt^{m-b}_{j-{\bR}+2g}$. Hence, we derive the association
\begin{equation}
\label{eqn:recurmap}
\sum_{g=0}^{\bR}\cbi{\bR}{g} \qt^{m-b}_{j-{\bR}+2g}
\sim 
(2z)^{\bR} \frac{\Bpol_{m-b}}{(2)^{(m-b)\HN-1}} .
\end{equation}
By \myeqref{eqn-scheme2}, the solution state $\wq{\mymat{W}}^{m+1}$ generates a pattern scaled by a factor of 1/2 with respect to the pattern corresponding to the solution state $\wq{\mymat{W}}^{m}$. Hence, a recurrence relation generating the correct pattern comprising any weighted average of $(2)^{L-m} \wq{\mymat{W}}^m$ over available values of $m$ will yield a valid solution state $\wq{\mymat{W}}^L$.

We define a \emph{ray} as any connection on a uniformly spaced graph which passes through nodes on every segment level $m$, $m=1,\,\cdots,\,M-1$. The sum of the recurrence weightings over any ray terminating at $m=L$ must be unity if the ray originates at the origin of a pattern flow at $m=0$, and zero otherwise. The coefficients of the Gaussian polynomials $\gauss{P}{k}$ ($k=1,\,\cdots,\,P$), denoted $\gauss{P}{k}^l$, possess the required properties. In~Fig.~\ref{fig:recur}, rays are shown summing to unity and zero, with a list of weightings satisfying these properties for all possible rays for the particular case of $N=2$.
Defining $P=2\HN+1$, the primitive form of the recurrence relation for $\wq{\mymat{W}}^L$ is
\beq
\label{eqn:primrecur}
\qt^{\wcheb{L}}_j
=
\sum_{k=1}^{\HN} \left(-1\right)^{k+1} \sum_{l=0}^{G_k}\gaussf{P}{k}^l
\left[\vph\left(\frac{1}{2}\right)^{k\HN}\qt^{\wcheb{L-k\HN}}_{j-\frac{1}{2}G_k+l}
-\left(\frac{1}{2}\right)^{(P-k)\HN}\qt^{\wcheb{L-(P-k)\HN}}_{j-\frac{1}{2}G_k+l}\right] 
+\left(\frac{1}{2}\right)^{P\HN}\qt^{\wcheb{L-P\HN}}_{j} ,
\eeq
\n where $G_k=kP-k^2$ is the degree of $\gauss{P}{k}$ for $k\le \HN$. We note that the Gaussian polynomial $\gauss{P}{k}$ possesses a unique representation as a summation of the binomial powers \mbox{$(1+q^2)^g$}, for \mbox{$g=0,\,\cdots,\,G_k/2$}, given by
\beq
\gaussf{P}{k}=\sum_{g=0}^{\frac{1}{2}G_k} \acoeff^{P,\,k}_g q^{\frac{1}{2}G_k-g}(1+q^2)^g ,
\label{eqn:gauss1}
\eeq 
where the coefficients $\acoeff^{P,\,k}_g$ follow the generating function
\beq 
\label{eqn:acoeffs}
\sum_{k=0}^{\infty}\sum_{g=0}^{\infty}(-1)^k(2)^g \acoeff^{P,\,k}_g(t)^k(z)^g=(1-t)\prod_{k=1}^{\HN}(1+(t)^2-2tC_{k}) .
\eeq
Then, using~\myeqref{eqn:gauss1}, we may recast \myeqref{eqn:primrecur} in the form
\beq
\label{eqn:primrecur2}
\qt^{\wcheb{L}}_j
=
\sum_{k=1}^{\HN} \left(-1\right)^{k+1}\sum_{g=0}^{\frac{1}{2}G_k}\acoeff^{P,\,k}_g\sum_{l=0}^g\binom{g}{l}
\left[\vph\left(\frac{1}{2}\right)^{k\HN}\qt^{\wcheb{L-k\HN}}_{j-g+2l}
-\left(\frac{1}{2}\right)^{(P-k)\HN}\qt^{\wcheb{L-(P-k)\HN}}_{j-g+2l}\right]
+\left(\frac{1}{2}\right)^{P\HN}\qt^{\wcheb{L-P\HN}}_{j} .
\eeq
Applying the association given in \myeqref{eqn:recurmap} to the terms in \myeqref{eqn:primrecur2}, the recurrence relation for $\Bpol_{M}$ follows as
\beq
\label{eqn:recur}
\Bpol_{M}
=
\sum_{k=1}^{\HN}\left(-1\right)^{k+1}  \left[\Bpol_{M-k} -\Bpol_{M-P+k} \right] \sum_{g=0}^{\frac{1}{2}G_k}\acoeff^{P,\,k}_g(2z)^{g} 
+\Bpol_{M-P}.
\eeq
We continue by noting that the generating function for $\Bpol_{k}$ derived from the recurrence relation given by \myeqref{eqn:recur} is 
\beq
\label{eqn:gen}
\sum_{k=0}^{\infty}(t)^k\Bpol_{k}
=
\frac{\sum_{k=0}^{2\HN}(t)^{k}b^{\HN}_k}
{1-\sum_{k=1}^{\HN}\left(-1\right)^{k+1} \left[(t)^{k} -(t)^{(P-k)} \right] \sum_{g=0}^{\frac{1}{2}G_k}\acoeff^{P,\,k}_g(2z)^{g} -(t)^{P}} ,
\eeq
\n where $b^{\HN}_k$ are coefficients determined by the seed states of $\Bpol_{m}$.
Appealing to \myeqref{eqn:acoeffs}, the generating function derived from the recurrence relation given by \myeqref{eqn:recur} is 
\beq
\label{eqn:gen2}
\sum_{k=0}^{\infty}(t)^k\Bpol_{k}
=
\frac{b^{\HN}_0}{1-t}+2\sum_{k=1}^{\HN}\frac{b^{\HN}_k(1-zt)}{1+(t)^2-2tC_{k}}
{(1-t)\prod_{k=1}^{\HN}(1+(t)^2-2tC_{k})} ,
\eeq
\n where $b^{\HN}_k$ are coefficients determined by the seed states of $\Bpol_{m}$. The normalization $\Bpol_{k}(1)=b^{\HN}_0(1)+\sum_{k=1}^{\HN}2b^{\HN}_k(1)$ has been imposed in order to fix the forms of the numerators in the separated fractions.

Noting that the generating function for $C_{km}$ is 
$\sum_{m=0}^{\infty}(t)^mC_{km}=(1-zt)/(1+(t)^2-2C_k)$, 
we conclude that 
$\Bpol_{k} =b^{\HN}_0+2\sum_{k=1}^{\HN}b^{\HN}_kC_{km}$. 
Consideration of the particular case $\HN=M=1$ indicates a correspondence between $b^{1}_k$ and $d^{1}_k$ is required in order to match the required solution pattern and normalization properties. A general correspondence between  $b^{\HN}_k$ and $d^{\HN}_k$ is established by considering successive values of $\HN$, with $M=1$, for the limiting case $d^{\HN}_k=0$, $0<k<\HN$.  This completes the derivation of the analytic expression for the \FRKC stability polynomial given by Eq.~\ref{eqn:FRKC}.

\section{Tests}
\label{sec:applic}

In this section numerical studies of two-dimensional two-species Brusselator diffusion-reaction problems are presented which confirm that high-order \FRKC stability polynomials meet all relevant linear order conditions and that the derived factorized numerical schemes are both stable and efficient. Split schemes, denoted \pFRKCs, are obtained by means of complex splitting techniques: linear diffusion operators are treated via \FRKC methods, while nonlinear reaction terms are integrated using standard Runge-Kutta techniques.  The performance of the second-order accurate unsplit \FRKCtwo scheme is compared to second-order \RKC and \CVODEtwo codes . Finally, comparisons are presented of  higher order split \FRKCs schemes (at orders 4 and 6), with fourth-order \ROCKf, and fifth-order \CVODE schemes.

\subsection{High-order splitting}
\label{sec:split}

\FRKC stability polynomials satisfy linear order conditions to an arbitrary order of accuracy. This property may be exploited in solving numerical problems for semi-linear stiff systems of equations through operator splitting methods~\citep{Castella09,hansenostermann09,Hansen01072010,dorsek2014high}. 
We note that in the literature, the linear and nonlinear terms of reaction-diffusion models have been decoupled under a variety of numerical integration techniques including: splitting methods~\citep[][and previous references]{mclachlan2002splitting}, Implicit-Explicit Runge-Kutta-Chebyshev (IMEX RKC) methods~\citep{ascher1997implicit,shampine2006irkc}, PIROCK~\citep{abdulle2013pirock}, and Local Linearization Runge-Kutta (LLRK) methods~\citep{cruz2006advanced,de2013local}. 

High-order splitting has been shown to give rise to an order reduction effect in some reaction-diffusion cases~\citep{warnez2013reduced}. For Dirichlet and Neumann boundary conditions, splitting techniques may result in order reduction at boundaries~\citep{hundsdorfer2003numerical,hansenostermann09}. It has also been observed that the full order is recovered on the interior of the computational domain when it is taken sufficiently far from the influence of the boundaries~\citep{lubich1995interior,lubich2013interior}. Boundary conditions for the separate operator updates are necessary to avoid order reduction effectively, however, as yet, no consistent treatment exists~\citep{hundsdorfer1995note}.

Assuming Eq.~\ref{eqn:autoneqn} is linearized and split in the form $\mymat{w}'=(\mymat{A}+\mymat{B})\mymat{w}$, the solution over a timestep $\TT$ requires an approximation to the operator ${\rm e}^{\TT(\mymat{A}+\mymat{B})}$. High-order approximations may be obtained through appropriate choice of partial steps $\TT_j$ where
\beq
\mymat{w}^{n+1}={\rm e}^{\TT_{k_J}  \mymat{B}}{\rm e}^{\TT_{k_{J-1}}  \mymat{A}}\cdots{\rm e}^{\TT_{k_3}  \mymat{B}}{\rm e}^{\TT_{k_2}  \mymat{A}}{\rm e}^{\TT_{k_1}  \mymat{B}} \mymat{w}^n .
\label{eqn:split}
\eeq
Formally, with support from numerical studies~\citep{Castella09,blanes2013optimized}, the splitting scheme given by Eq.~\ref{eqn:split} may be may be extended to the semi-linear parabolic form of Eq.~\ref{eqn:autoneqn} given by 
\beq
w'=Aw+f_B(w)
\label{eqn:semi}
\eeq
by replacing the exponential operator ${\rm e}^{\TT_{k_j}  \mymat{B}}$ with a step of the nonlinear equation $w'=f_B(w)$ over the interval $\TT_{k_j}$. For reference, the complex splitting schemes used in this work are provided in Table~\ref{tab:splitting}. 

\subsection{Brusselator}
\label{sec:bruss}

The Brusselator~\citep{lefever1971chemical,hairernonstiff} is a stiff nonlinear diffusion-reaction problem describing chemical kinetics of a tri-molecular chemical reaction.
The test case considered here is a two-dimensional hybrid of the one- and two-dimensional Brusselator problems presented by \citet{hairernonstiff}, and \citet{hairerstiff}, with governing equations given by

\begin{eqnarray}
\label{eqn:bruss}
\plone{v}{t}&=&\epsilon\left(\pltwo{v}{x_1} +\pltwo{v}{x_2}\right) +A -(B+1)v +w v^2 ,\nonumber\\
\plone{w}{t}&=&\epsilon\left(\pltwo{w}{x_1} +\pltwo{w}{x_2}\right)  +B v  -v^2w ,
\label{eqn:bruss}
\end{eqnarray}

\n and initial conditions $v(0,\,x)=A+\sin(2\pi x)$, $v(0,\,x)=B/A+\cos(2\pi y)$. The initial state is therefore a simple perturbation of the equilibrium solution. The problem is configured with parameters  $\epsilon=0.02$, $A=1$, and $B=3$, and the solution is obtained at $t=2$, or $t=8$, on the domain $0\le x_1\le 1$, $0\le x_2\le 1$, under periodic boundary conditions.

\begin{figure*}
  \centering
  \begin{minipage}{160mm}
    \centering\leavevmode
    {
      \resizebox{79mm}{!}{\input{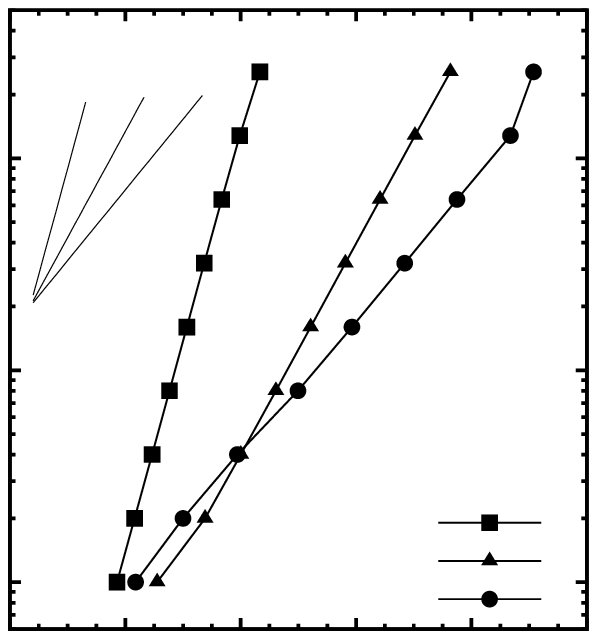}}
    }
    \end{minipage}
    \caption{${\rm L}_1$ errors plotted against number of timesteps, $N_T$, for species $v$ of the two-dimensional Brusselator problem. Results correspond to the split problem, with linear diffusion treated via \FRKC methods at orders 2, 4, and 6, and nonlinear reaction terms integrated via standard techniques. Guide lines are shown for $({\rm L}_1 {\rm error})^{-1/2}$, $({\rm L}_1 {\rm error})^{-1/4}$, $({\rm L}_1 {\rm error})^{-1/6}$. Table~\ref{tab:tests} gives the values for all points shown. 
}
  \label{fig:spliterr}       
\end{figure*}

\begin{table}
\caption{Error convergence results for the two-dimensional Brusselator test problem solved via split \FRKCs schemes with $N=2,\,4,\,6$.
Each row corresponds to a specific test with columns listing: $N$, order of accuracy; $N_{T}$, the number of timesteps; ${\rm L}_1$ norm of the error between the approximate and exact solutions; ${\rm L}_1$ order of convergence with reference to previous row; ${\rm L}_\infty$ error; ${\rm L}_\infty$ order. Errors refer to the solution for species $v$.
${\rm L}_1$ errors are also shown in Fig.\ref{fig:spliterr}.
}
\label{tab:tests}       
\begin{tabular}{SSS[table-format=+1.2e+2,round-precision=3,round-mode=figures,scientific-notation=false]S[table-format=+1.2,round-precision=3,round-mode=figures,scientific-notation=false]S[table-format=+1.2e+2,round-precision=3,round-mode=figures,scientific-notation=false]S[table-format=+1.2,round-precision=3,round-mode=figures,scientific-notation=false]}
\hline\noalign{\smallskip}
\mbox{$N$}  & \mbox{$N_{T}$} &  \mbox{${\rm L}_1$ error} & \mbox{${\rm L}_1$ order} & \mbox{${\rm L}_\infty$ error} & \mbox{${\rm L}_\infty$ order} \\
\noalign{\smallskip}\hline\noalign{\smallskip}
2 & 50  &  \num{1.923047255799590000e-04} & \shm & \num{4.248482044044850000e-04} &  \shm \\
 & 100  &  \num{4.755240946551430000e-05} & 2.015803866077385336e+00 & \num{1.031805089500920000e-04} &  2.041776999551262136e+00 \\
 & 200  &  \num{1.182896298891790000e-05} & 2.007194841531917949e+00 & \num{2.545600802639090000e-05} &  2.019092367102712651e+00 \\
 & 400  &  \num{2.949392749514790000e-06} & 2.003833748632897380e+00 & \num{6.321266152439280000e-06} &  2.009720730747864613e+00 \\
 & 800  &  \num{7.359924467439030000e-07} & 2.002655083009724606e+00 & \num{1.574405088160020000e-06} &  2.005406771512227309e+00 \\
 & 1600  &  \num{1.833582863596900000e-07} & 2.005025494389472197e+00 & \num{3.918623382722720000e-07} &  2.006387960406966391e+00 \\
 & 3200  &  \num{4.529472363220850000e-08} & 2.017250559958480981e+00 & \num{9.676913599425410000e-08} &  2.017728036778103179e+00 \\
 & 6400  &  \num{1.078285056551180000e-08} & 2.070604380114969277e+00 & \num{2.303118871614320000e-08} &  2.070958105551529665e+00 \\
 & 12800  &  \num{2.156483902820590000e-09} & 2.321985768705283762e+00 & \num{4.605647757160820000e-09} &  2.322112893097319542e+00 \\
\noalign{\smallskip}\hline\noalign{\smallskip}
4  & 50  &  \num{7.849476548709050000e-06} & \shm & \num{1.214296053266650000e-05} &  \shm \\
 & 100  &  \num{1.712653734392720000e-07} & 5.518291048589189403e+00 & \num{2.760541458361530000e-07} &  5.459025124702944322e+00 \\
 & 200  &  \num{9.956434090902570000e-09} & 4.104460553722142553e+00 & \num{1.649630151895990000e-08} &  4.064736755260232589e+00 \\
 & 400  &  \num{6.053701907067760000e-10} & 4.039739590137725320e+00 & \num{1.017323911418380000e-09} &  4.019291602452332116e+00 \\
 & 800  &  \num{3.759071837546550000e-11} & 4.009369245243659542e+00 & \num{6.359471976802310000e-11} &  3.999728305444845164e+00 \\
 & 1600  &  \num{2.346410582997070000e-12} & 4.001849097881972391e+00 & \num{3.982018599406330000e-12} &  3.997335119569081621e+00 \\
 & 3200  &  \num{1.465934649063320000e-13} & 4.000562788102071956e+00 & \num{2.491352393482750000e-13} &  3.998498954160067909e+00 \\
 & 6400  &  \num{9.129292516544790000e-15} & 4.005173918280449566e+00 & \num{1.553087086020310000e-14} &  4.003718467282063037e+00 \\
 & 12800  &  \num{5.369128101193310000e-16} & 4.087743330019177728e+00 & \num{9.145245324915230000e-16} &  4.085973046418977373e+00 \\
\noalign{\smallskip}\hline\noalign{\smallskip}
6  & 50  &  \num{4.414315191987980000e-05} & \shm & \num{7.369227683744250000e-05} &  \shm \\
 & 100  &  \num{9.918522664336970000e-07} & 5.475920583423954594e+00 & \num{2.354579044159230000e-06} &  4.967972368241269584e+00 \\
 & 200  &  \num{1.308610464215580000e-08} & 6.244017633916991968e+00 & \num{2.879421243882340000e-08} &  6.353546482187166839e+00 \\
 & 400  &  \num{1.039341118097720000e-10} & 6.976222668768329002e+00 & \num{1.780414757681160000e-10} &  7.337421688353259208e+00 \\
 & 800  &  \num{1.385427643918060000e-12} & 6.229194057226279228e+00 & \num{2.340304382578150000e-12} &  6.249373373557355822e+00 \\
 & 1600  &  \num{2.056271001892120000e-14} & 6.074157141595060685e+00 & \num{3.524155793577230000e-14} &  6.053274666912389511e+00 \\
 & 3200  &  \num{3.163774194159640000e-16} & 6.022239968759627936e+00 & \num{5.485520891690410000e-16} &  6.005505271225686924e+00 \\
 & 6400  &  \num{4.436553311154090000e-18} & 6.156063520280490875e+00 & \num{9.866239769618090000e-18} &  5.796984463540586602e+00 \\
 & 12800  &  \num{7.005677629828620000e-19} & 2.662842796154909993e+00 & \num{2.005774019098180000e-18} &  2.298341274569748739e+00 \\
\noalign{\smallskip}\hline
\end{tabular}
\end{table}

\subsection{Linear order conditions}
\label{sec:linsplit}

The semi-discrete form of Eq.~\ref{eqn:bruss} may be written $\mymat{w}'=\mymat{A}\mymat{w}+\mymat{f}_B(\mymat{w})$, where $\mymat{A}$ describes the discretization of the Laplacian with respect to $x_1$ and $x_2$, and $\mymat{f}_B(\mymat{w})$ contains the reaction terms. Linear diffusion terms are integrated using \FRKC methods and nonlinear reaction terms via standard techniques. The linear order properties of the \FRKC stability polynomials are confirmed by considering the convergence rates of the approximated solution to the exact solution at $t=2$ as a function of step size.

For all presented results, we use $M=20$, and the approximation $\dmm_M= 1$. The number of grid points is 400 in each of the two spatial variables. For these parameters, the \FRKC stability polynomials achieve approximately $81\%$ ($\beta_R=1066.667$), $74\%$ ($\beta_R=1600$), and $73\%$ ($\beta_R=2133.333$) of the optimal intervals for $N=2,\,4,\,6$ respectively. With respect to the corresponding standard explicit Runge-Kutta schemes, these values represent a speedup in efficiency by factors of approximately $27$ for $N=2$, and $30$ for both $N=4$ and $N=6$. All polynomials are damped with damping parameter $\nu_0=0.05$, reducing the stability domains' real extents by factors of $1-\nu_0/N$. Finally, in order to meet the specified solution time, timesteps are scaled by 0.9846, 0.9001, 0.7563 for $N=2,\,4,\,6$ respectively. Quadruple precision is used in all calculations.  Results are presented in Table~\ref{tab:tests} where the ${\rm L}_1$ and ${\rm L}_\infty$ errors  are shown over a range of resolutions at each considered value of $N$. Fig.~\ref{fig:spliterr} illustrates the dependence of the ${\rm L}_1$ errors on the number of timesteps, $N_T$, for species $v$. With the exception of the final point for the sixth-order integration, where machine precision is exceeded, all solutions are converging in good agreement with the nominal orders of accuracy (i.e. $({\rm error})^{-1/N}$). Fitting the ${\rm L}_1$ errors yields observed orders $2.04\pm 0.01$, $4.08\pm 0.04$, $6.1\pm 0.2$ for $N=2,\,4,\,6$ respectively, while the ${\rm L}_\infty$ errors give $2.05\pm 0.01$, $4.09\pm 0.05$, and $6.0\pm 0.2$. We conclude that \FRKC methods demonstrate internal stability and comply with linear order conditions to the specified order of accuracy.


\begin{table}
\vspace*{-4cm}
\caption{Errors from \pFRKCtwo, \pRKC, \CVODEtwo from tests of the two-dimensional Brusselator problem. The number of timesteps,  $N_{T}$, and the number of stages per timestep, $L$, (or the error tolerance, Tol, in the case of \pCVODE) are given in the first two columns respectively. The wall time taken for each run,  $\Twall$, is presented in the third column. ${\rm L}_1$ and ${\rm L}_\infty$ errors for both species are presented in the remaining columns. The ${\rm L}_1$ error for species $v$ is plotted in Fig.~\ref{fig:timing}~(a). 
}
\label{tab:allschemes-nonlinbruss2d}       
\begin{tabular}{S[table-format=4]S[table-format=2, retain-unity-mantissa = false]S[table-format=3]S[table-format=+1.2e+2,round-precision=3,round-mode=figures,scientific-notation=false]S[table-format=+1.2e+2,round-precision=3,round-mode=figures,scientific-notation=false]S[table-format=+1.2e+2,round-precision=3,round-mode=figures,scientific-notation=false]S[table-format=+1.2e+2,round-precision=3,round-mode=figures,scientific-notation=false]}
\hline\noalign{\smallskip}
 &   &  & \multicolumn{2}{c}{Species $v$}  & \multicolumn{2}{c}{Species $w$}  \\
\mbox{$N_{T}$} & \mbox{$L$/Tol}  & \mbox{$\Twall$ (s)} &\mbox{${\rm L}_1$ error}&\mbox{${\rm L}_\infty$ error}&\mbox{${\rm L}_1$ error}&\mbox{${\rm L}_\infty$ error}\\
\noalign{\smallskip}\hline\noalign{\smallskip}
 \multicolumn{7}{c}{\FRKCtwo} \\
\noalign{\smallskip}\hline\noalign{\smallskip}
50 & 80 & 25 & 5.15787943803883e-03 & 5.52105627241439e-03 & 1.29783368482789e-03 & 1.40325921187068e-03 \\ 
137 & 48 & 42 & 1.40665113988448e-03 & 1.45067474895111e-03 & 6.06462259572033e-04 & 6.24623089919396e-04 \\ 
308 & 32 & 63 & 3.24197192606298e-04 & 3.33867744283678e-04 & 1.64041704052819e-04 & 1.68390167734955e-04 \\ 
548 & 24 & 84 & 1.07635233708955e-04 & 1.10810120237836e-04 & 5.66253118736774e-05 & 5.80856503673566e-05 \\ 
1317 & 16 & 137 & 1.84501069012491e-05 & 1.89824914853531e-05 & 1.00190327587211e-05 & 1.02662594154701e-05 \\ 
2341 & 12 & 186 & 6.02840372015595e-06 & 6.20165855780286e-06 & 3.29890118854421e-06 & 3.37962861696184e-06 \\ 
5266 & 8 & 288 & 1.26252591662206e-06 & 1.29877749022178e-06 & 6.94156377073120e-07 & 7.11101084460708e-07 \\ 
9361 & 6 & 399 & 4.22548567317491e-07 & 4.34769230617249e-07 & 2.32062466637704e-07 & 2.37798035929160e-07 \\ 
21062 & 4 & 637 & 9.48007979837573e-08 & 9.76127694229945e-08 & 5.16316916257378e-08 & 5.29656263292821e-08 \\ 
105026 & 2 & 1881 & 5.71082662725181e-09 & 5.88018012059877e-09 & 3.11257306120083e-09 & 3.19294279904625e-09 \\ 
\noalign{\smallskip}\hline\noalign{\smallskip}
\multicolumn{7}{c}{\RKC} \\
\noalign{\smallskip}\hline\noalign{\smallskip}
39 & 90 & 32 & 4.96798864290415e-03 & 5.81159116726382e-03 & 8.05272896395659e-03 & 8.54181086070249e-03 \\ 
78 & 64 & 45 & 6.58865105550828e-04 & 7.64019955872497e-04 & 6.49719676277658e-04 & 7.02313780106323e-04 \\ 
137 & 48 & 60 & 2.49433256748253e-04 & 2.82644719558522e-04 & 1.07532447217680e-04 & 1.25841259463355e-04 \\ 
309 & 32 & 90 & 5.06628170383671e-05 & 5.73147878752955e-05 & 1.35631155899566e-05 & 1.75073170634032e-05 \\ 
549 & 24 & 119 & 1.59318320884719e-05 & 1.80565396552534e-05 & 3.83646821945689e-06 & 5.11641262135321e-06 \\ 
1237 & 16 & 177 & 3.12815973215841e-06 & 3.55248236472150e-06 & 7.27335452114715e-07 & 9.85355877780592e-07 \\ 
2206 & 12 & 237 & 9.91793622840052e-07 & 1.12728076873125e-06 & 2.27506512368714e-07 & 3.10200417130702e-07 \\ 
5007 & 8 & 359 & 1.99877564265932e-07 & 2.27246223172273e-07 & 4.49630652310351e-08 & 6.17051802986879e-08 \\ 
9012 & 6 & 486 & 6.56744239399709e-08 & 7.45743156116419e-08 & 1.39743115580349e-08 & 1.94187339541685e-08 \\ 
21028 & 4 & 750 & 1.52299736352904e-08 & 1.71106844248925e-08 & 1.88486211322586e-09 & 3.02968827803340e-09 \\ 
39428 & 3 & 1056 & 6.56064075613694e-09 & 7.19742998533945e-09 & 4.12710340581823e-10 & 7.97714339029199e-10 \\ 
\noalign{\smallskip}\hline\noalign{\smallskip}
\multicolumn{7}{c}{\CVODEtwo} \\
\noalign{\smallskip}\hline\noalign{\smallskip}
1226 & \num{5e-06} & 107 & 2.36652711225613e-03 & 2.42335011774042e-03 & 9.64715442290514e-04 & 1.01187037455719e-03 \\ 
1499 & \num{1e-06} & 128 & 7.78863720620782e-04 & 8.25869242740529e-04 & 5.85467050810005e-05 & 8.44605518561803e-05 \\ 
2534 & \num{1e-07} & 194 & 4.69349048533187e-05 & 5.83050380713601e-05 & 2.99298226620009e-05 & 3.90658380913234e-05 \\ 
4991 & \num{1e-08} & 312 & 3.05520822643671e-05 & 3.22055805415111e-05 & 3.13497232432506e-05 & 3.26552356688659e-05 \\ 
10029 & \num{1e-09} & 523 & 5.46840756238039e-06 & 5.59896656793235e-06 & 4.35227523310348e-06 & 4.45454773045917e-06 \\ 
22763 & \num{1e-10} & 987 & 7.05393524382444e-07 & 7.14043161709199e-07 & 5.05636002258471e-07 & 5.07658319826021e-07 \\ 
48444 & \num{1e-11} & 1706 & 1.55242083496003e-07 & 1.57912152465300e-07 & 1.07284639767243e-07 & 1.08852774571844e-07 \\ 
109474 & \num{1e-12} & 3405 & 2.96092238937090e-08 & 3.01112046408036e-08 & 2.06026057046033e-08 & 2.09580124366227e-08 \\ 
232430 & \num{1e-13} & 6756 & 6.05180899861807e-09 & 6.16217743498737e-09 & 4.15989999569127e-09 & 4.23947810190839e-09 \\ 
\noalign{\smallskip}\hline
\end{tabular}
\end{table}


\subsection{Second-order comparative studies}
\label{secondcomp}

Since all order conditions are linear at second-order, \FRKCtwo schemes will naturally maintain second-order accuracy for nonlinear problems without the necessity of splitting or composition methods. Here we present comparative studies between \FRKCtwo and a number of alternative numerical integration methods. In particular, we provide comparisons with the \RKC method~\citep{sommeijer1998rkc} which, similarly to \pFRKCtwo, depends on the properties of Chebyshev polynomials. We also compare results with a GMRES Krylov-preconditioned BDF integrator from the \CVODE numerical integration package~\citep{cohen1996cvode}.  The \CVODE solver maintains a specified tolerance by means of adaptive stepping up to a maximum fifth-order accuracy.  However, the order is restricted to 2 for the \CVODEtwo solver used in these comparisons.

We proceed by considering the two-dimensional Brusselator problem described in Section~\ref{sec:bruss} with the solution taken at time $t=8$. The stepsize is fixed for individual tests of the explicit schemes and the number of internal stages is optimized for the selected stepsize. As such, each of the numerical solutions generated for these tests is derived from a single distinct stability polynomial. In general, however, error control procedures may be implemented~\citep{sommeijer1998rkc} which will result in stability polynomials of varied degree contributing to particular solutions.  The optimal efficiency for extended stability explicit solvers follows $\Twall \propto ({\rm error})^{-1/2N}$ (where $\Twall$ is the wall-time required for computation of a particular solution).

\begin{figure*}
  \centering
  \begin{minipage}{160mm}
    \centering\leavevmode
    {
      \hspace*{-5mm}
      \resizebox{79mm}{!}{\input{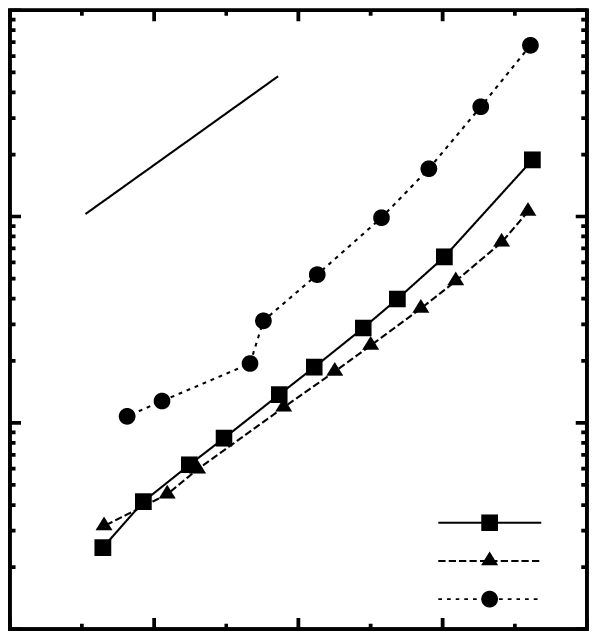}}
      \hspace*{-5mm}
      \resizebox{79mm}{!}{\input{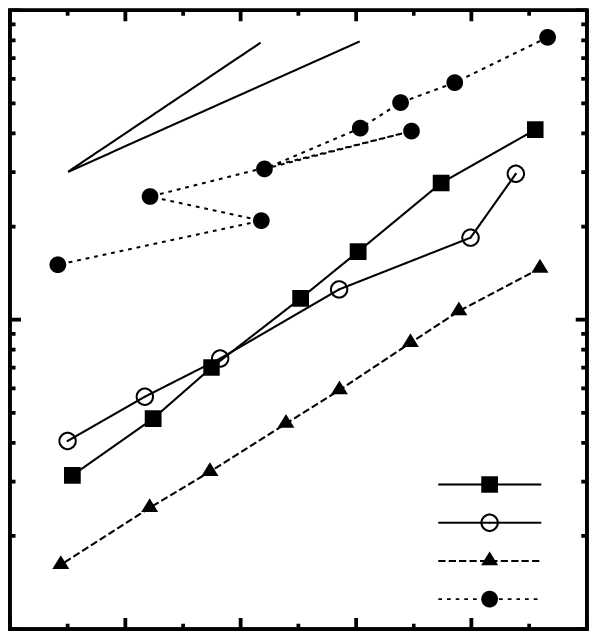}}    
      }
  \end{minipage}
  \caption{Performance results derived from the ${\rm L}_1$ error in the solution to species $v$ in the stiff nonlinear Brusselator problem. Panel (a) shows results for the second-order schemes \pFRKCtwo, \pRKC, and \CVODEtwo (see Table~\ref{tab:allschemes-nonlinbruss2d}). A guide line is shown for $({\rm L}_1 {\rm error})^{-1/4}$. Panel (b) shows data for the higher-order schemes \pFRKCfs, \pFRKCxs, \pROCKf, and \CVODE (see Table~\ref{tab:allschemes-nonlinbruss2d-fourth}). Guide lines are shown for $({\rm L}_1 {\rm error})^{-1/8}$ and $({\rm L}_1 {\rm error})^{-1/12}$. 
  }
  \label{fig:timing}       
\end{figure*}

Results are provided in Table~\ref{tab:allschemes-nonlinbruss2d} for \pFRKCtwo, \pRKC, and \pCVODEtwo. The ${\rm L}_1$ errors for species $v$ are plotted in Fig.~\ref{fig:timing}~(a) against the time required for the simulations to be carried out on a standard desktop machine at double precision. While the \FRKCtwo solver requires complex arithmetic, this is compensated by smaller errors than for the \RKC solver at equivalent numbers of timesteps. Overall, \FRKCtwo runs at about 70\% of the efficiency of \pRKC. As previously noted, following a similar strategy to \citet{lebedev2000explicit,lebedev1994solve} will improve performance.

\begin{table}
  \vspace*{-4cm}
  \caption{Errors from \pFRKCfs, \pFRKCxs, \pROCKf, \CVODE from tests of the two-dimensional Brusselator problem. The number of timesteps,  $N_{T}$, and the number of stages per timestep, $L$, (or the error tolerance, Tol, in the case of \pCVODE) are given in the first two columns respectively. The wall time taken for each run,  $\Twall$, is presented in the third column. ${\rm L}_1$ and ${\rm L}_\infty$ errors for both species are presented in the remaining columns. The ${\rm L}_1$ error for species $v$ is plotted in Fig.~\ref{fig:timing}~(b).
}
\label{tab:allschemes-nonlinbruss2d-fourth}       
\begin{tabular}{S[table-format=4]S[table-format=2, retain-unity-mantissa = false]S[table-format=3]S[table-format=+1.2e+2,round-precision=3,round-mode=figures,scientific-notation=false]S[table-format=+1.2e+2,round-precision=3,round-mode=figures,scientific-notation=false]S[table-format=+1.2e+2,round-precision=3,round-mode=figures,scientific-notation=false]S[table-format=+1.2e+2,round-precision=3,round-mode=figures,scientific-notation=false]}
\hline\noalign{\smallskip}
 &   &  & \multicolumn{2}{c}{Species $v$}  & \multicolumn{2}{c}{Species $w$}  \\
\mbox{$N_{T}$} & \mbox{$L$/Tol}  & \mbox{$\Twall$ (s)} &\mbox{${\rm L}_1$ error}&\mbox{${\rm L}_\infty$ error}&\mbox{${\rm L}_1$ error}&\mbox{${\rm L}_\infty$ error}\\
\noalign{\smallskip}\hline\noalign{\smallskip}
\multicolumn{7}{c}{\FRKCfs} \\
\noalign{\smallskip}\hline\noalign{\smallskip}
23 & 96 & 31 & 8.26691002948547e-04 & 8.49306219867563e-04 & 5.48731637876371e-04 & 5.61898538177230e-04 \\ 
51 & 64 & 48 & 3.29744212337992e-05 & 3.30095833092070e-05 & 2.34753380488157e-05 & 2.36504957635830e-05 \\ 
91 & 48 & 70 & 3.21377528647666e-06 & 3.24982679233976e-06 & 2.30909017683216e-06 & 2.31919005133641e-06 \\ 
222 & 32 & 117 & 9.16985421946903e-08 & 9.27145891171932e-08 & 6.65612561332962e-08 & 6.68418558280592e-08 \\ 
395 & 24 & 166 & 9.17282148660825e-09 & 9.27390653160387e-09 & 6.67745031668054e-09 & 6.70516242529118e-09 \\ 
887 & 16 & 277 & 3.33943184804841e-10 & 3.37594174837363e-10 & 2.41943504573905e-10 & 2.42937447936242e-10 \\ 
1577 & 12 & 411 & 7.82937042836940e-12 & 7.90634224756559e-12 & 4.03344859734034e-12 & 4.08983957811415e-12 \\ 
\noalign{\smallskip}\hline\noalign{\smallskip}
\multicolumn{7}{c}{\FRKCxs} \\
\noalign{\smallskip}\hline\noalign{\smallskip}
5 & 144 & 41 & 1.00382477250832e-03 & 1.01889280441481e-03 & 8.60002205246307e-04 & 8.73928909087329e-04 \\ 
10 & 96 & 56 & 4.58163656559490e-05 & 4.61783678507022e-05 & 3.57166360874907e-05 & 3.61128620713913e-05 \\ 
17 & 72 & 75 & 2.26213497530911e-06 & 2.31455295929273e-06 & 1.78921598498283e-06 & 1.84300512984237e-06 \\ 
42 & 48 & 125 & 1.96872763007908e-08 & 2.00544141293335e-08 & 1.54843316297698e-08 & 1.56641348780795e-08 \\ 
75 & 36 & 184 & 1.03664006594062e-10 & 1.04228625730229e-10 & 8.27408076609082e-11 & 8.33120239462914e-11 \\ 
168 & 24 & 296 & 1.69706761837807e-11 & 1.71864744658023e-11 & 1.34006446175783e-11 & 1.34818822772331e-11 \\ 
\noalign{\smallskip}\hline\noalign{\smallskip}
\multicolumn{7}{c}{\ROCKf} \\
\noalign{\smallskip}\hline\noalign{\smallskip}
56 & 102 & 16 & 1.31040875788998e-03 & 7.26101509592514e-03 & 1.34026147807672e-03 & 7.18769370125028e-03 \\ 
130 & 67 & 25 & 3.78905809662608e-05 & 2.13340888322922e-04 & 2.88033589448704e-05 & 1.49214446136137e-04 \\ 
224 & 51 & 32 & 3.40220594294174e-06 & 1.83413956738043e-05 & 2.76387935403856e-06 & 1.42375845391118e-05 \\ 
451 & 36 & 46 & 1.64097827024340e-07 & 8.17753980975056e-07 & 1.37983334656462e-07 & 7.55826718989994e-07 \\ 
749 & 28 & 59 & 1.94628779004127e-08 & 8.89507125734212e-08 & 1.59783075083963e-08 & 8.13930165488586e-08 \\ 
1483 & 20 & 84 & 1.14322871749978e-09 & 4.63730231992088e-09 & 8.98741926153024e-10 & 4.73977723736141e-09 \\ 
2345 & 16 & 107 & 1.65319930206520e-10 & 5.58347812429361e-10 & 1.24927683514686e-10 & 5.66223956610656e-10 \\ 
4282 & 12 & 146 & 6.47976926004823e-12 & 3.32340821529442e-11 & 6.13379610436837e-12 & 2.79898326738248e-11 \\ 
\noalign{\smallskip}\hline\noalign{\smallskip}
\multicolumn{7}{c}{\CVODE} \\
\noalign{\smallskip}\hline\noalign{\smallskip}
2001 & \num{1e-06} & 151 & 1.47968992972911e-03 & 1.50988455130130e-03 & 1.13248723032444e-03 & 1.15281142996304e-03 \\ 
3203 & \num{1e-07} & 209 & 4.38198872701961e-07 & 1.26104644948555e-06 & 4.83379804245847e-07 & 1.70671587440552e-06 \\ 
3309 & \num{1e-08} & 250 & 3.73529304208553e-05 & 3.89677841834413e-05 & 2.90865904452726e-05 & 3.03218126569416e-05 \\ 
6628 & \num{1e-09} & 407 & 1.09263646063812e-09 & 5.14157960651573e-09 & 9.32578304652098e-10 & 4.06277100850616e-09 \\ 
4818 & \num{1e-10} & 307 & 3.87921214937521e-07 & 3.88761884906330e-07 & 3.27229083783301e-07 & 3.28106620095525e-07 \\ 
6246 & \num{1e-11} & 416 & 8.46169152288190e-09 & 8.97000695943007e-09 & 4.87701704104532e-09 & 4.94559904140601e-09 \\ 
8376 & \num{1e-12} & 503 & 1.69031802263603e-09 & 1.76191083944843e-09 & 1.31819641830261e-09 & 1.37274125400211e-09 \\ 
10515 & \num{1e-13} & 584 & 1.95629389247720e-10 & 1.96573646249476e-10 & 1.38114625167218e-10 & 1.40751632571323e-10 \\ 
16036 & \num{1e-14} & 817 & 4.79214444515374e-12 & 5.29287724759797e-12 & 3.37686754808830e-12 & 3.84381415585722e-12 \\ 
\noalign{\smallskip}\hline
\end{tabular}
\end{table}


\subsection{High-order comparative studies}

An advantage of \FRKC methods over other extended stability methods is extensibility to arbitrarily high-order linear stability polynomials. In order to apply high- (above second-) order \FRKC stability polynomials to nonlinear problems, complex splitting techniques may be employed (as demonstrated in Sec.~\ref{sec:linsplit}). In the following tests, we consider fourth- and sixth-order solutions of the two-dimensional Brusselator problem. We note that finishing stages based on the theory of the composition of Butcher series may also be may be used to meet nonlinear order conditions~\citep{hairernonstiff,medovikov1998high,abdulle2002fourth}. While composition methods may, in principle, offer improved efficiency over splitting techniques, in the case of \pROCKf, the application of finishing stages has been observed to result in order reduction problems, as well as erratic convergence properties, limiting the number of internal stages to a relatively small number of internal stages\cite{abdulle2001chebyshev, hundsdorfer2003numerical}. (The limit adopted within the \ROCKf  code is $L=152$.) Furthermore, the number of nonlinear order conditions, and hence the complexity of the composition strategy, grows rapidly with increasing order~\cite{zbMATH05292586}: there are four nonlinear order conditions at fourth-order, 31 order conditions at sixth-order, and 192 at eighth-order.

We present comparisons of the split schemes \FRKCfs and \FRKCxs with the fourth-order \ROCKf scheme. Reference solutions obtained using the \CVODE solver are also presented for integrations carried out to a maximum fifth-order accuracy. The test conditions are otherwise as described in Sec.~\ref{secondcomp}. All data are presented in Table~\ref{tab:allschemes-nonlinbruss2d-fourth}, with ${\rm L}_1$ errors for species $v$ plotted in Fig.~\ref{fig:timing}~(b) against the time required for the simulations at double precision. We note that \FRKCfs is shown to run at approximately half the efficiency of \pROCKf. This is primarily due to the additional computational overhead of carrying out calculations with complex values quantities, which, as noted, may be countered to some degree by rolling conjugate pair calculations together using the scheme presented by~\citet{lebedev2000explicit,lebedev1994solve}. In terms of efficiency, for the presented problem, the \FRKCxs and \FRKCfs methods lie approximately midway between \ROCKf and \pCVODE. Except at the very lowest acceleration parameters considered, the \FRKCs trials show the predicted behavior (i.e. $\Twall \propto ({\rm L}_1 {\rm error})^{-1/2N}$).

\section{Conclusions}
\label{sec:conc}

The fully prescribed analytic form of a new class of extended stability polynomials which satisfy all required linear order conditions to arbitrarily high-order has been presented. Factorized Runge-Kutta-Chebyshev (\pFRKC) stability polynomials are derived from first principles by inductive considerations of the implied recurrence relations. At order $N$, the \FRKC polynomial of rank $N$, and degree $L=MN$, is shown to have the form of a summation of Chebyshev polynomials, with degrees at intervals of $M$, up to degree $L$. The $N+1$ weightings of the contributing Chebyshev polynomials are chosen to comply with the $N$ linear order conditions, coupled with a constraint. A damping procedure for broadening the stability domain of the \FRKC stability polynomials to a finite width along the real axis is described which preserves the order of accuracy. The resultant stability polynomials have been demonstrated to have $81\%$, $74\%$ and $73\%$ of the optimal intervals for orders $2,\,4,\,6$ respectively.  \FRKC numerical integration schemes are represented as a sequence of $L$ sequenced forward Euler steps (stages) involving complex-valued timesteps constructed from the roots of \FRKC stability polynomials of degree $L$. Internal stability is maintained by means of a sequencing algorithm, which limits the maximum internal amplification factor to $\sim L^2$: reserving 8 digits for accuracy, a hypothetical scheme of 10,000 stages is therefore viable in a numerical integration carried out at 16 digit precision.

Split \FRKCs schemes have been applied at orders 2, 4, and 6, to the linear diffusion operator in numerical experiments on a stiff two-dimensional Brusselator reaction-diffusion system leading to the verification of expected convergence rates, and hence compliance with the necessary linear order conditions.

We have presented comparative studies of the performance of \FRKCtwo with \pRKC, an established explicit extended stability code, and \pCVODEtwo, an implicit preconditioned BDF solver from the \CVODE suite limited to second-order accuracy. \FRKCtwo has been shown to be substantially more efficient than the \CVODEtwo solver, while performing at about 70\% of the efficiency of \pRKC.

At higher orders, nonlinear order conditions require special attention. We have considered treatment of these nonlinear conditions through complex splitting techniques in efficiency tests of higher order (4 and 6)  split \FRKCs schemes in comparison with results from the the fourth-order \ROCKf code, which uses composition methods, and the implicit fifth-order \CVODE solver. The tested \FRKCs methods are found to have intermediate efficiency to \ROCKf and \CVODE. We propose implementing conjugate pairing and Butcher series composition methods in future high-order implementations of \FRKC methods.

\section{Acknowledgments}

Calculations involving the coefficients and roots of polynomials are handled with the \url{gmp} and \url{mpfr} libraries.  Polynomial roots are obtained by means of the \url{MPSolve} package.
MPSolve was written by Dario Andrea Bini and Giuseppe Fiorentino, Dipartimento di Matematica, Universita' di Pisa, Italy.

The author thanks an anonymous referee for constructive comments which have contributed to significant improvements in this paper.

\section*{References}

\bibliography{poly23}

\begin{thebibliography}{54}
\expandafter\ifx\csname natexlab\endcsname\relax\def\natexlab#1{#1}\fi
\providecommand{\url}[1]{\texttt{#1}}
\providecommand{\href}[2]{#2}
\providecommand{\path}[1]{#1}
\providecommand{\DOIprefix}{doi:}
\providecommand{\ArXivprefix}{arXiv:}
\providecommand{\URLprefix}{URL: }
\providecommand{\Pubmedprefix}{pmid:}
\providecommand{\doi}[1]{\href{http://dx.doi.org/#1}{\path{#1}}}
\providecommand{\Pubmed}[1]{\href{pmid:#1}{\path{#1}}}
\providecommand{\bibinfo}[2]{#2}
\ifx\xfnm\relax \def\xfnm[#1]{\unskip,\space#1}\fi
\bibitem[{Markov(1890)}]{markov1890question}
\bibinfo{author}{A.~A. Markov},
\newblock \bibinfo{title}{On a question by {DI} {Mendeleev}},
\newblock \bibinfo{journal}{Zapiski Imperatorskoi Akademii Nauk}
  \bibinfo{volume}{62} (\bibinfo{year}{1890}) \bibinfo{pages}{1--24}.
\bibitem[{Markov(1892)}]{markov1892functions}
\bibinfo{author}{V.~Markov},
\newblock \bibinfo{title}{On functions deviating least from zero in a given
  interval},
\newblock \bibinfo{journal}{Izdat. Imp. Akad. Nauk, St. Petersburg}
  (\bibinfo{year}{1892}) \bibinfo{pages}{218--258}.
\bibitem[{Saul’ev(1960)}]{saul1960integration}
\bibinfo{author}{V.~Saul’ev},
\newblock \bibinfo{title}{Integration of parabolic equations by the grid
  method},
\newblock \bibinfo{journal}{Fizmatgiz, Moscow}  (\bibinfo{year}{1960}).
\bibitem[{Guillou and Lago(1960)}]{guillou1960domaine}
\bibinfo{author}{A.~Guillou}, \bibinfo{author}{B.~Lago},
\newblock \bibinfo{title}{Domaine de stabilit{\'e} associ{\'e} aux formules
  d’int{\'e}gration num{\'e}rique d’{\'e}quations diff{\'e}rentielles, a
  pas s{\'e}par{\'e}s et a pas li{\'e}s. recherche de formules a grand rayon de
  stabilit{\'e}},
\newblock \bibinfo{journal}{Ier Congr. Ass. Fran. Calcul., AFCAL}
  (\bibinfo{year}{1960}) \bibinfo{pages}{43--56}.
\bibitem[{Gentzsch and Schluter(1978)}]{gentzsch1978one}
\bibinfo{author}{W.~Gentzsch}, \bibinfo{author}{A.~Schluter},
\newblock \bibinfo{title}{On one-step methods with cyclic stepsize changes for
  solving parabolic differential equations},
\newblock \bibinfo{journal}{Z. Angew. Math. Mech} \bibinfo{volume}{58}
  (\bibinfo{year}{1978}) \bibinfo{pages}{T415--T416}.
\bibitem[{van~der Houwen(1996)}]{vanderHouwen1996261}
\bibinfo{author}{P.~van~der Houwen},
\newblock \bibinfo{title}{The development of {Runge-Kutta} methods for partial
  differential equations},
\newblock \bibinfo{journal}{Applied Numerical Mathematics} \bibinfo{volume}{20}
  (\bibinfo{year}{1996}) \bibinfo{pages}{261 -- 272}.
\bibitem[{Alexiades et~al.(1996)Alexiades, Amiez, and
  Gremaud}]{alexiades1996super}
\bibinfo{author}{V.~Alexiades}, \bibinfo{author}{G.~Amiez},
  \bibinfo{author}{P.~Gremaud},
\newblock \bibinfo{title}{Super-time-stepping acceleration of explicit schemes
  for parabolic problems},
\newblock \bibinfo{journal}{Communications in numerical methods in engineering}
  \bibinfo{volume}{12} (\bibinfo{year}{1996}) \bibinfo{pages}{31--42}.
\bibitem[{O'Sullivan and Downes(2006)}]{o2006explicit}
\bibinfo{author}{S.~O'Sullivan}, \bibinfo{author}{T.~P. Downes},
\newblock \bibinfo{title}{An explicit scheme for multifluid
  magnetohydrodynamics},
\newblock \bibinfo{journal}{Monthly Notices of the Royal Astronomical Society}
  \bibinfo{volume}{366} (\bibinfo{year}{2006}) \bibinfo{pages}{1329--1336}.
\bibitem[{O'Sullivan and Downes(2007)}]{o2007three}
\bibinfo{author}{S.~O'Sullivan}, \bibinfo{author}{T.~P. Downes},
\newblock \bibinfo{title}{A three-dimensional numerical method for modelling
  weakly ionized plasmas},
\newblock \bibinfo{journal}{Monthly Notices of the Royal Astronomical Society}
  \bibinfo{volume}{376} (\bibinfo{year}{2007}) \bibinfo{pages}{1648--1658}.
\bibitem[{O'Sullivan and O'Sullivan(2011)}]{o2011acceleration}
\bibinfo{author}{S.~O'Sullivan}, \bibinfo{author}{C.~O'Sullivan},
\newblock \bibinfo{title}{On the acceleration of explicit finite difference
  methods for option pricing},
\newblock \bibinfo{journal}{Quantitative Finance} \bibinfo{volume}{11}
  (\bibinfo{year}{2011}) \bibinfo{pages}{1177--1191}.
\bibitem[{Lebedev(2000)}]{lebedev2000explicit}
\bibinfo{author}{V.~I. Lebedev},
\newblock \bibinfo{title}{Explicit difference schemes for solving stiff
  problems with a complex or separable spectrum},
\newblock \bibinfo{journal}{Computational mathematics and mathematical physics}
  \bibinfo{volume}{40} (\bibinfo{year}{2000}) \bibinfo{pages}{1729--1740}.
\bibitem[{Medovikov(1998)}]{medovikov1998high}
\bibinfo{author}{A.~A. Medovikov},
\newblock \bibinfo{title}{High order explicit methods for parabolic equations},
\newblock \bibinfo{journal}{BIT Numerical Mathematics} \bibinfo{volume}{38}
  (\bibinfo{year}{1998}) \bibinfo{pages}{372--390}.
\bibitem[{van Der~Houwen and Sommeijer(1980)}]{vanderhouwen1980}
\bibinfo{author}{P.~J. van Der~Houwen}, \bibinfo{author}{B.~P. Sommeijer},
\newblock \bibinfo{title}{On the internal stability of explicit, m-stage
  {Runge-Kutta} methods for large m-values},
\newblock \bibinfo{journal}{ZAMM - Journal of Applied Mathematics and Mechanics
  / Zeitschrift für Angewandte Mathematik und Mechanik} \bibinfo{volume}{60}
  (\bibinfo{year}{1980}) \bibinfo{pages}{479--485}.
\bibitem[{Verwer(1996)}]{verwer1996explicit}
\bibinfo{author}{J.~G. Verwer},
\newblock \bibinfo{title}{Explicit {Runge-Kutta} methods for parabolic partial
  differential equations},
\newblock \bibinfo{journal}{Applied Numerical Mathematics} \bibinfo{volume}{22}
  (\bibinfo{year}{1996}) \bibinfo{pages}{359--379}.
\bibitem[{Sommeijer et~al.(1998)Sommeijer, Shampine, and
  Verwer}]{sommeijer1998rkc}
\bibinfo{author}{B.~Sommeijer}, \bibinfo{author}{L.~Shampine},
  \bibinfo{author}{J.~Verwer},
\newblock \bibinfo{title}{{RKC:} an explicit solver for parabolic {PDEs}},
\newblock \bibinfo{journal}{Journal of Computational and Applied Mathematics}
  \bibinfo{volume}{88} (\bibinfo{year}{1998}) \bibinfo{pages}{315--326}.
\bibitem[{Abdulle and Medovikov(2001)}]{abdulle2001second}
\bibinfo{author}{A.~Abdulle}, \bibinfo{author}{A.~A. Medovikov},
\newblock \bibinfo{title}{Second order {Chebyshev} methods based on orthogonal
  polynomials},
\newblock \bibinfo{journal}{Numerische Mathematik} \bibinfo{volume}{90}
  (\bibinfo{year}{2001}) \bibinfo{pages}{1--18}.
\bibitem[{Abdulle(2002)}]{abdulle2002fourth}
\bibinfo{author}{A.~Abdulle},
\newblock \bibinfo{title}{Fourth order {Chebyshev} methods with recurrence
  relation},
\newblock \bibinfo{journal}{SIAM Journal on Scientific Computing}
  \bibinfo{volume}{23} (\bibinfo{year}{2002}) \bibinfo{pages}{2041--2054}.
\bibitem[{Martin-Vaquero and Janssen(2009)}]{martin2009second}
\bibinfo{author}{J.~Martin-Vaquero}, \bibinfo{author}{B.~Janssen},
\newblock \bibinfo{title}{Second-order stabilized explicit {Runge-Kutta}
  methods for stiff problems},
\newblock \bibinfo{journal}{Computer Physics Communications}
  \bibinfo{volume}{180} (\bibinfo{year}{2009}) \bibinfo{pages}{1802--1810}.
\bibitem[{Meyer et~al.(2012)Meyer, Balsara, and Aslam}]{meyer2012second}
\bibinfo{author}{C.~D. Meyer}, \bibinfo{author}{D.~S. Balsara},
  \bibinfo{author}{T.~D. Aslam},
\newblock \bibinfo{title}{A second-order accurate {Super} {TimeStepping}
  formulation for anisotropic thermal conduction},
\newblock \bibinfo{journal}{Monthly Notices of the Royal Astronomical Society}
  \bibinfo{volume}{422} (\bibinfo{year}{2012}) \bibinfo{pages}{2102--2115}.
\bibitem[{Meyer et~al.(2014)Meyer, Balsara, and Aslam}]{meyer2014stabilized}
\bibinfo{author}{C.~D. Meyer}, \bibinfo{author}{D.~S. Balsara},
  \bibinfo{author}{T.~D. Aslam},
\newblock \bibinfo{title}{A stabilized {Runge-Kutta-Legendre} method for
  explicit super-time-stepping of parabolic and mixed equations},
\newblock \bibinfo{journal}{Journal of Computational Physics}
  \bibinfo{volume}{257} (\bibinfo{year}{2014}) \bibinfo{pages}{594--626}.
\bibitem[{Hairer et~al.(1993)Hairer, N{\o}rsett, and Wanner}]{hairernonstiff}
\bibinfo{author}{E.~Hairer}, \bibinfo{author}{S.~N{\o}rsett},
  \bibinfo{author}{G.~Wanner},
\newblock \bibinfo{title}{Solving ordinary differential equations {I}: Nonstiff
  problems},
\newblock \bibinfo{journal}{Springer series in computational mathematics}
  \bibinfo{volume}{8} (\bibinfo{year}{1993}).
\bibitem[{{Butcher}(2008)}]{zbMATH05292586}
\bibinfo{author}{J.~C. {Butcher}}, \bibinfo{title}{Numerical methods for
  ordinary differential equations. 2nd revised ed.}, \bibinfo{edition}{2nd
  revised ed.} ed., \bibinfo{publisher}{Hoboken, NJ: John Wiley \& Sons},
  \bibinfo{year}{2008}. \DOIprefix\doi{10.1002/9780470753767}.
\bibitem[{Van Der~Houwen(1977)}]{van1977}
\bibinfo{author}{P.~J. Van Der~Houwen}, \bibinfo{title}{Construction of
  integration formulas for initial value problems}, \bibinfo{publisher}{North
  Holland}, \bibinfo{year}{1977}.
\bibitem[{Abdulle(2001)}]{abdulle2001chebyshev}
\bibinfo{author}{A.~Abdulle}, \bibinfo{title}{Chebyshev methods based on
  orthogonal polynomials}, Ph.D. thesis, \bibinfo{year}{2001}.
\bibitem[{Bakker(1971)}]{bakker71}
\bibinfo{author}{M.~Bakker}, \bibinfo{title}{Analytisch {Aspekten} {Van} {Een}
  {Minimaxprobleem}}, \bibinfo{type}{CWI Technical Report}
  \bibinfo{number}{Stichting Mathematisch Centrum. Toegepaste Wiskunde-TN
  62/71}, \bibinfo{year}{1971}. \URLprefix
  \url{http://oai.cwi.nl/oai/asset/7896/7896A.pdf}.
\bibitem[{Lomax(1968)}]{lomax1968construction}
\bibinfo{author}{H.~Lomax},
\newblock \bibinfo{title}{On the construction of highly stable, explicit
  numerical methods for integrating coupled {ODEs} with parasitic eigenvalues},
\newblock \bibinfo{journal}{NASA Technical Note NASAIN D/4547}
  (\bibinfo{year}{1968}).
\bibitem[{Riha(1972)}]{riha72}
\bibinfo{author}{W.~Riha},
\newblock \bibinfo{title}{Optimal stability polynomials},
\newblock \bibinfo{journal}{Computing} \bibinfo{volume}{9}
  (\bibinfo{year}{1972}) \bibinfo{pages}{37--43}.
\bibitem[{Lebedev(1994)}]{lebedev1994solve}
\bibinfo{author}{V.~Lebedev},
\newblock \bibinfo{title}{How to solve stiff systems of differential equations
  by explicit methods},
\newblock \bibinfo{journal}{Numerical methods and applications}
  (\bibinfo{year}{1994}) \bibinfo{pages}{45--80}.
\bibitem[{Lebedev and Finogenov(1976)}]{lebedev1976utilization}
\bibinfo{author}{V.~Lebedev}, \bibinfo{author}{S.~Finogenov},
\newblock \bibinfo{title}{Utilization of ordered {Chebyshev} parameters in
  iterative methods},
\newblock \bibinfo{journal}{USSR Computational Mathematics and Mathematical
  Physics} \bibinfo{volume}{16} (\bibinfo{year}{1976}) \bibinfo{pages}{70--83}.
\bibitem[{Verwer et~al.(1990)Verwer, Hundsdorfer, and
  Sommeijer}]{verwer1990convergence}
\bibinfo{author}{J.~Verwer}, \bibinfo{author}{W.~Hundsdorfer},
  \bibinfo{author}{B.~Sommeijer},
\newblock \bibinfo{title}{Convergence properties of the {Runge-Kutta-Chebyshev}
  method},
\newblock \bibinfo{journal}{Numerische Mathematik} \bibinfo{volume}{57}
  (\bibinfo{year}{1990}) \bibinfo{pages}{157--178}.
\bibitem[{Hairer and Wanner(1996)}]{hairerstiff}
\bibinfo{author}{E.~Hairer}, \bibinfo{author}{G.~Wanner},
\newblock \bibinfo{title}{Solving ordinary differential equations {II}: Stiff
  and differential-algebraic problems},
\newblock \bibinfo{journal}{Springer series in computational mathematics}
  \bibinfo{volume}{14} (\bibinfo{year}{1996}).
\bibitem[{Ketcheson et~al.(2013)Ketcheson, L{\'o}czi, and
  Parsani}]{ketcheson2013propagation}
\bibinfo{author}{D.~I. Ketcheson}, \bibinfo{author}{L.~L{\'o}czi},
  \bibinfo{author}{M.~Parsani},
\newblock \bibinfo{title}{Propagation of internal errors in explicit
  {Runge-Kutta} methods and internal stability of ssp and extrapolation
  methods},
\newblock \bibinfo{journal}{arXiv preprint arXiv:1309.1317}
  (\bibinfo{year}{2013}).
\bibitem[{Lebedev and Finogenov(1973)}]{lebedev1973solution}
\bibinfo{author}{V.~I. Lebedev}, \bibinfo{author}{S.~Finogenov},
\newblock \bibinfo{title}{Solution of the problem of parameter ordering in
  {Chebyshev} iteration methods},
\newblock \bibinfo{journal}{Zhurnal Vychislitel'noi Matematiki i
  Matematicheskoi Fiziki} \bibinfo{volume}{13} (\bibinfo{year}{1973})
  \bibinfo{pages}{18--33}.
\bibitem[{Marchuk and Lebedev(1986)}]{marchuk1986numerical}
\bibinfo{author}{G.~Marchuk}, \bibinfo{author}{V.~I. Lebedev},
  \bibinfo{title}{Numerical methods in the theory of neutron transport},
  \bibinfo{publisher}{Harwood Academic Pub., New York, NY},
  \bibinfo{year}{1986}.
\bibitem[{Hundsdorfer and Verwer(2003)}]{hundsdorfer2003numerical}
\bibinfo{author}{W.~Hundsdorfer}, \bibinfo{author}{J.~G. Verwer},
  \bibinfo{title}{Numerical solution of time-dependent
  advection-diffusion-reaction equations}, volume~\bibinfo{volume}{33},
  \bibinfo{publisher}{Springer}, \bibinfo{year}{2003}.
\bibitem[{Mart{\'\i}n-Vaquero et~al.(2014)Mart{\'\i}n-Vaquero, Khaliq, and
  Kleefeld}]{martin2014stabilized}
\bibinfo{author}{J.~Mart{\'\i}n-Vaquero}, \bibinfo{author}{A.~Khaliq},
  \bibinfo{author}{B.~Kleefeld},
\newblock \bibinfo{title}{Stabilized explicit {Runge-Kutta} methods for
  multi-asset {American} options},
\newblock \bibinfo{journal}{Computers \& Mathematics with Applications}
  \bibinfo{volume}{67} (\bibinfo{year}{2014}) \bibinfo{pages}{1293--1308}.
\bibitem[{Castella et~al.(2009)Castella, Chartier, Descombes, and
  Vilmart}]{Castella09}
\bibinfo{author}{F.~Castella}, \bibinfo{author}{P.~Chartier},
  \bibinfo{author}{S.~Descombes}, \bibinfo{author}{G.~Vilmart},
\newblock \bibinfo{title}{Splitting methods with complex times for parabolic
  equations},
\newblock \bibinfo{journal}{BIT Numerical Mathematics} \bibinfo{volume}{49}
  (\bibinfo{year}{2009}) \bibinfo{pages}{487--508}.
\bibitem[{Hansen and Ostermann(2009)}]{hansenostermann09}
\bibinfo{author}{E.~Hansen}, \bibinfo{author}{A.~Ostermann},
\newblock \bibinfo{title}{High order splitting methods for analytic semigroups
  exist},
\newblock \bibinfo{journal}{BIT Numerical Mathematics} \bibinfo{volume}{49}
  (\bibinfo{year}{2009}) \bibinfo{pages}{527--542}.
\bibitem[{Hansen and Ostermann(2010)}]{Hansen01072010}
\bibinfo{author}{E.~Hansen}, \bibinfo{author}{A.~Ostermann},
\newblock \bibinfo{title}{Dimension splitting for quasilinear parabolic
  equations},
\newblock \bibinfo{journal}{IMA Journal of Numerical Analysis}
  \bibinfo{volume}{30} (\bibinfo{year}{2010}) \bibinfo{pages}{857--869}.
\bibitem[{D{\"o}rsek and Hansen(2014)}]{dorsek2014high}
\bibinfo{author}{P.~D{\"o}rsek}, \bibinfo{author}{E.~Hansen},
\newblock \bibinfo{title}{High order splitting schemes with complex timesteps
  and their application in mathematical finance},
\newblock \bibinfo{journal}{Journal of Computational and Applied Mathematics}
  \bibinfo{volume}{262} (\bibinfo{year}{2014}) \bibinfo{pages}{234--243}.
\bibitem[{McLachlan and Quispel(2002)}]{mclachlan2002splitting}
\bibinfo{author}{R.~I. McLachlan}, \bibinfo{author}{G.~R.~W. Quispel},
\newblock \bibinfo{title}{Splitting methods},
\newblock \bibinfo{journal}{Acta Numerica} \bibinfo{volume}{11}
  (\bibinfo{year}{2002}) \bibinfo{pages}{341--434}.
\bibitem[{Ascher et~al.(1997)Ascher, Ruuth, and Spiteri}]{ascher1997implicit}
\bibinfo{author}{U.~M. Ascher}, \bibinfo{author}{S.~J. Ruuth},
  \bibinfo{author}{R.~J. Spiteri},
\newblock \bibinfo{title}{Implicit-explicit {Runge-Kutta} methods for
  time-dependent partial differential equations},
\newblock \bibinfo{journal}{Applied Numerical Mathematics} \bibinfo{volume}{25}
  (\bibinfo{year}{1997}) \bibinfo{pages}{151--167}.
\bibitem[{Shampine et~al.(2006)Shampine, Sommeijer, and
  Verwer}]{shampine2006irkc}
\bibinfo{author}{L.~Shampine}, \bibinfo{author}{B.~Sommeijer},
  \bibinfo{author}{J.~Verwer},
\newblock \bibinfo{title}{{IRKC:} an {IMEX} solver for stiff diffusion-reaction
  {PDEs}},
\newblock \bibinfo{journal}{Journal of computational and applied mathematics}
  \bibinfo{volume}{196} (\bibinfo{year}{2006}) \bibinfo{pages}{485--497}.
\bibitem[{Abdulle and Vilmart(2013)}]{abdulle2013pirock}
\bibinfo{author}{A.~Abdulle}, \bibinfo{author}{G.~Vilmart},
\newblock \bibinfo{title}{{PIROCK:} a swiss-knife partitioned implicit-explicit
  orthogonal {Runge-Kutta} {Chebyshev} integrator for stiff
  diffusion-advection-reaction problems with or without noise},
\newblock \bibinfo{journal}{Journal of Computational Physics}
  \bibinfo{volume}{242} (\bibinfo{year}{2013}) \bibinfo{pages}{869--888}.
\bibitem[{Cruz et~al.(2006)Cruz, Biscay, Carbonell, Jimenez, and
  Ozaki}]{cruz2006advanced}
\bibinfo{author}{H.~D.~l. Cruz}, \bibinfo{author}{R.~Biscay},
  \bibinfo{author}{F.~Carbonell}, \bibinfo{author}{J.~Jimenez},
  \bibinfo{author}{T.~Ozaki},
\newblock \bibinfo{title}{Advanced numerical algorithms-{Local} {Linearization}
  {Runge-Kutta} (llrk) methods for solving ordinary differential equations},
\newblock \bibinfo{journal}{Lecture Notes in Computer Science}
  \bibinfo{volume}{3991} (\bibinfo{year}{2006}) \bibinfo{pages}{132--139}.
\bibitem[{De~la Cruz et~al.(2013)De~la Cruz, Biscay, Jim{\'e}nez, and
  Carbonell}]{de2013local}
\bibinfo{author}{H.~De~la Cruz}, \bibinfo{author}{R.~Biscay},
  \bibinfo{author}{J.~C. Jim{\'e}nez}, \bibinfo{author}{F.~Carbonell},
\newblock \bibinfo{title}{{Local} {Linearization} {Runge-Kutta} methods: A
  class of {A-stable} explicit integrators for dynamical systems},
\newblock \bibinfo{journal}{Mathematical and Computer Modelling}
  \bibinfo{volume}{57} (\bibinfo{year}{2013}) \bibinfo{pages}{720--740}.
\bibitem[{Warnez and Muite(2013)}]{warnez2013reduced}
\bibinfo{author}{M.~Warnez}, \bibinfo{author}{B.~Muite},
\newblock \bibinfo{title}{Reduced temporal convergence rates in high-order
  splitting schemes},
\newblock \bibinfo{journal}{arXiv preprint arXiv:1310.3901}
  (\bibinfo{year}{2013}).
\bibitem[{Lubich and Ostermann(1995)}]{lubich1995interior}
\bibinfo{author}{C.~Lubich}, \bibinfo{author}{A.~Ostermann},
\newblock \bibinfo{title}{Interior estimates for time discretizations of
  parabolic equations},
\newblock \bibinfo{journal}{Applied numerical mathematics} \bibinfo{volume}{18}
  (\bibinfo{year}{1995}) \bibinfo{pages}{241--251}.
\bibitem[{Lubich and Makridakis(2013)}]{lubich2013interior}
\bibinfo{author}{C.~Lubich}, \bibinfo{author}{C.~Makridakis},
\newblock \bibinfo{title}{Interior a posteriori error estimates for time
  discrete approximations of parabolic problems},
\newblock \bibinfo{journal}{Numerische Mathematik} \bibinfo{volume}{124}
  (\bibinfo{year}{2013}) \bibinfo{pages}{541--557}.
\bibitem[{Hundsdorfer and Verwer(1995)}]{hundsdorfer1995note}
\bibinfo{author}{W.~Hundsdorfer}, \bibinfo{author}{J.~Verwer},
\newblock \bibinfo{title}{A note on splitting errors for advection-reaction
  equations},
\newblock \bibinfo{journal}{Applied Numerical Mathematics} \bibinfo{volume}{18}
  (\bibinfo{year}{1995}) \bibinfo{pages}{191--199}.
\bibitem[{Blanes et~al.(2013)Blanes, Casas, Chartier, and
  Murua}]{blanes2013optimized}
\bibinfo{author}{S.~Blanes}, \bibinfo{author}{F.~Casas},
  \bibinfo{author}{P.~Chartier}, \bibinfo{author}{A.~Murua},
\newblock \bibinfo{title}{Optimized high-order splitting methods for some
  classes of parabolic equations},
\newblock \bibinfo{journal}{Mathematics of Computation} \bibinfo{volume}{82}
  (\bibinfo{year}{2013}) \bibinfo{pages}{1559--1576}.
\bibitem[{Lefever and Nicolis(1971)}]{lefever1971chemical}
\bibinfo{author}{R.~Lefever}, \bibinfo{author}{G.~Nicolis},
\newblock \bibinfo{title}{Chemical instabilities and sustained oscillations},
\newblock \bibinfo{journal}{Journal of theoretical Biology}
  \bibinfo{volume}{30} (\bibinfo{year}{1971}) \bibinfo{pages}{267--284}.
\bibitem[{Cohen and Hindmarsh(1996)}]{cohen1996cvode}
\bibinfo{author}{S.~D. Cohen}, \bibinfo{author}{A.~C. Hindmarsh},
\newblock \bibinfo{title}{{CVODE,} a stiff/nonstiff {ODE} solver in {C}},
\newblock \bibinfo{journal}{Computers in physics} \bibinfo{volume}{10}
  (\bibinfo{year}{1996}) \bibinfo{pages}{138--143}.
\bibitem[{Strang(1968)}]{strang1968construction}
\bibinfo{author}{G.~Strang},
\newblock \bibinfo{title}{On the construction and comparison of difference
  schemes},
\newblock \bibinfo{journal}{SIAM Journal on Numerical Analysis}
  \bibinfo{volume}{5} (\bibinfo{year}{1968}) \bibinfo{pages}{506--517}.

\end{thebibliography}

\newpage
\appendix
\section{Scheme patterns $M=20$}
\label{app:patterns}

\begin{eqnarray}
d^{2}_0 =&& \frac{267}{400} \nonumber\\
d^{2}_1 =&-&\frac{1}{1800} \nonumber\\
d^{2}_2 =&& \frac{1201}{7200} 
\end{eqnarray}

\begin{eqnarray}
d^{4}_0 =&& \frac{3126039467}{6144000000} \nonumber\\
d^{4}_1 =&& \frac{244573733}{7680000000} \nonumber\\
d^{4}_2 =&& \frac{3212226667}{15360000000} \nonumber\\
d^{4}_3 =&-& \frac{63194381}{7680000000} \nonumber\\
d^{4}_4 =&& \frac{789861181}{61440000000} 
\end{eqnarray}

\begin{eqnarray}
d^{6}_0 =&& \frac{7446093942631413209}{17915904000000000000} \nonumber\\
d^{6}_1 =&& \frac{158532158867283313}{2985984000000000000} \nonumber\\
d^{6}_2 =&& \frac{1022936325403301087}{4777574400000000000} \nonumber\\
d^{6}_3 =&-& \frac{35821864811075087}{10749542400000000000} \nonumber\\
d^{6}_4 =&& \frac{1048968349471238687}{35831808000000000000} \nonumber\\
d^{6}_5 =&-& \frac{32100268736824717}{17915904000000000000} \nonumber\\
d^{6}_6 =&& \frac{180240686854539517}{214990848000000000000} 
\end{eqnarray}

\newpage
\section{Splitting schemes}
\label{app:splitting}
%
\begin{longtable}{SSS[table-format=2.42,table-text-alignment=left]}
\caption{Complex operator splitting parameters for $N=2,\,4,\,6$~\cite{strang1968construction,Castella09,blanes2013optimized}. The final row for each quoted value of $N$ lists: $J$, the number of distinct sweep configurations required; $k_1\,\cdots\,k_J$, the sequence of $J$ sweeps, labeled by $j$, required for a single extended interval to order-$N$.
The remaining rows are in pairs listing: $j$, the index of the distinct sweep; $\Re(T_j)$, the real component of the sweep timescale; $\Im(T_j)$, the second row lists the imaginary part of the sweep timescale.\\
}
\label{tab:splitting}\\
\hline\noalign{\smallskip}\\
\mbox{$N$}   & \mbox{$j$} & \mbox{$\Re(T_j)$}   \\  
  &  &  \secrow \mbox{$\Im(T_j)$}   \\  
\noalign{\smallskip}\cline{2-3}\noalign{\smallskip}
  & \mbox{$J$} &\mbox{$k_1\,\cdots\,k_J$}   \\ 
\noalign{\smallskip}\hline\noalign{\smallskip}
2  & 1 & 1.0 \\
  & & \secrow 0.0\\
 &   2 & 0.5 \\
 &  & \secrow 0.0\\
\noalign{\smallskip}\cline{2-3}\noalign{\smallskip}
 &  3 & \multicolumn{1}{l}{\mbox{\hspace*{1ex} 2\sq 1\sq 2}}\\
\noalign{\smallskip}\hline\noalign{\smallskip}
4   & 1 &      {$1/4$}   \\
 &  & \secrow  {0}     \\
 &   2 &       {1/10}  \\
 &  & \secrow  {\hspace{-1.5ex}$-$1/30} \\
 &   3 &       {4/15}  \\ 
 &  & \secrow  {2/15}  \\
 &   4 &       {4/15}  \\
 &  & \secrow  {\hspace{-1.5ex}$-$1/5}  \\
\noalign{\smallskip}\cline{2-3}\noalign{\smallskip}
 &  9 & \multicolumn{1}{l}{\mbox{\hspace*{1ex} 2\sq 1\sq 3\sq 1\sq 4\sq 1\sq 3\sq 1\sq 2}}\\
\noalign{\smallskip}\hline\noalign{\smallskip}
6   & 1 & 0.0625 \\
 &  & \secrow 0.0\\
 &   2 & 0.02469487608701806464091086499684224783860 \\
 &  & \secrow  -0.00787479556290687705817157794952694216320 \\
 &   3 & 0.06381347402130269977936630418820014696320 \\
 &  & \secrow 0.03536576103414332780462940464971474181270 \\
 &   4 & 0.06842509403031644197039700782174468405850 \\
 &  & \secrow -0.06226224445074867699533254064444759604610 \\
 &   5 & 0.08804770109226783762699719586940866757720 \\
 &  & \secrow 0.04547387150229870438376254918797742644469 \\
 &   6 & 0.02368961112984706069614191247000936432533 \\
 &  & \secrow 0.00962432606408962405769803529063730666395 \\
 &   7 & 0.04272972238677338220296430057707421855388 \\
 &  & \secrow -0.03399440392395761055408394845784435826499 \\
 &   8 & 0.12233468631684577296042851700196256307880 \\
 &  & \secrow -0.01043585907975251066938082710059054955178 \\
 &   9 & 0.04189843282969388604353685060726223976426 \\
 &  & \secrow 0.06936249263169638427515817430714426213030 \\
 &   10 & 0.04873280421186970815851409293499173568080 \\
 &  & \secrow -0.09051829642972473048855853856612858205130 \\
\noalign{\smallskip}\cline{2-3}\noalign{\smallskip}
 &  33 & \multicolumn{1}{l}{\mbox{\hspace*{1ex} 2\sq 1\sq 3\sq 1\sq 4\sq 1\sq 5\sq 1\sq 6\sq 1\sq 7\sq 1\sq 8\sq 1\sq 9\sq 1\sq 10\sq 1\sq 9\sq 1\sq 8\sq 1\sq 7\sq 1\sq 6\sq 1\sq 5\sq 1\sq 4\sq 1\sq 3\sq 1\sq 2}}\\
\noalign{\smallskip}\hline
\end{longtable}

\end{document}